\documentclass[aps,pre,reprint,10pt]{revtex4-2}
\usepackage[english]{babel}
\usepackage{amsmath,amssymb,amsthm,mathtools,bm}
\usepackage{euscript,mathrsfs}
\usepackage{graphicx}
\usepackage{wrapfig}
\usepackage[dvipsnames]{xcolor}
\usepackage{indentfirst}
\usepackage{amsbsy}
\usepackage{wasysym}
\usepackage{cancel}
\usepackage{verbatim}
\usepackage{empheq}
\usepackage{adjustbox}
\usepackage[pdfencoding=auto]{hyperref}
\definecolor{urlcolor}{HTML}{005F5F}
\definecolor{linkcolor}{HTML}{005F5F}
\hypersetup{pdfstartview=FitH,  linkcolor=linkcolor,urlcolor=blue, colorlinks=true,citecolor=blue}
\usepackage{layouts}

\usepackage[dvipsnames]{xcolor}
\usepackage{tikz-cd}
\usepackage{tikz}
\usetikzlibrary{positioning}
\usetikzlibrary{calc}

\definecolor{mylightred}{RGB}{211,79,73}
\definecolor{mydarkred}{RGB}{199,44,38}
\definecolor{mylightgreen}{RGB}{78,153,67}
\definecolor{mydarkgreen}{RGB}{43,129,33}
\definecolor{mylightpurple}{RGB}{150,107,178}
\definecolor{mydarkpurple}{RGB}{126,78,160}
\definecolor{mylightblue}{RGB}{49,101,205}
\definecolor{mydarkblue}{RGB}{20,92,205}

\tikzset{
  juliadot/.style args={#1,#2}{shape=circle,line width=0.03ex,minimum width=0.4ex,fill=#1,draw=#2}
}

\newcommand\julialetter[1]{{\strut\fontfamily{cmss}\bfseries\selectfont{#1}}}

\DeclareRobustCommand\julia{%
\begin{tikzpicture}[baseline=0mm, every node/.style={inner sep=0mm, outer sep=0mm}]
\node[anchor=base]        (j) at (0,0) {\julialetter{\j}};
\node[anchor=base, right=0ex of j] (u) {\julialetter{u}};
\node[anchor=base, right=0ex of u] (l) {\julialetter{l}};
\node[anchor=base, right=0ex of l] (i) {\julialetter{\i}};
\node[anchor=base, right=0ex of i] (a) {\julialetter{a}};
\path let \p1 = (j) in node[juliadot={mylightblue,mydarkblue}] (bluedot) at (\x1+0.02ex,1.4ex) {};
\path let \p1 = (i) in node[juliadot={mylightred,mydarkred}] (reddot) at (\x1,1.4ex) {};
\path let \p1 = (reddot) in node[juliadot={mylightpurple,mydarkpurple}] (purpledot) at (\x1+0.5ex,\y1) {};
\path let \p1 = (reddot) in node[juliadot={mylightgreen,mydarkgreen}] (greendot) at (\x1+0.25ex,\y1+0.42ex) {};
\end{tikzpicture}%
}

\DeclareMathOperator{\im}{\mathrm{Im}\,}
\DeclareMathOperator{\re}{\mathrm{Re}\,}
\DeclareMathOperator{\tr}{\mathrm{tr}}
\DeclareMathOperator{\sgn}{\mathrm{sgn}}

\DeclareMathOperator{\arccosh}{\mathrm{arccosh}}

\begin{document}

\title{Duck hunting with quantum mechanics}
\author{Artem Alexandrov}
\email{alexandrov.aa@mi-ras.ru}
\affiliation{Steklov Mathematical Institute of Russian Academy of Sciences, Moscow 119333, Russia}
\affiliation{Moscow Institute of Physics and Technology, Dolgoprudny, 141700, Russia}

\begin{abstract}
    We bridge two sides of singular perturbation theory: the classical theory of slow-fast systems and the semi-classical approach to quantum mechanical systems. For a specific but physically important class of dynamical systems, we show that purely classical and exotic objects, so-called canard solutions, are shadows of instantons in the corresponding quantum system. We demonstrate that canard solutions exist in a domain of parameter space whose boundaries are determined by an instanton action. We illustrate our statements analytically for the relevant example, the overdamped Josephson junction, and confirm them numerically. For the Josephson junction, the canard window is the exponentially narrow gap between consecutive Shapiro steps.
\end{abstract}

\maketitle

\section{Introduction}

Singular perturbation theory is known by different names in modern science: the WKB method in quantum mechanics~\cite{kawai2005algebraic}, the large deviation theory in probability~\cite{varadhan1966asymptotic,touchette2009largedeviation,dembo2010largedeviations, freidlin2012random}, the boundary layer theory in fluid dynamics~\cite{prandtl1904boundarylayer, schlichting2017boundarylayer}, the slow-fast theory in dynamical systems~\cite{tikhonov1952systems, fenichel1979gspt, jones1995gspt, kuehn2015multiscale}, and the renormalization group approach~\cite{chen1994rg, chen1996rg}. It provides a unifying asymptotic framework, but sometimes this common mathematical foundation eludes due to the specific jargon of each field of science and the massive flow of technical mathematical proofs and theorems. In this paper, we demonstrate a direct connection between the existence of canard trajectories in slow-fast systems on a torus with exact quantization in the corresponding quantum mechanical system.

The slow-fast dynamics can demonstrate very special trajectories, which are called \emph{canards}. The existence of canard trajectories was shown in~\cite{benoit1981chasse, benoit1983systemes, dumortier1996cycles} (it was start for the duck hunting, \emph{chasse au canard}) and then this unusual behavior was observed in different dynamical systems~\cite{desroches2011curvature, moehlis2006hodgkinhuxley, brons1991belousov}. From the physical point of view, canards are well-known in nonlinear physics: they are related to mixed-mode oscillations~\cite{harvey2011calcium, rotstein2008stellate, desroches2012mmo, zhan2018pituitary, liu2020pyramidal}, bursting and spiking regimes in neuronal models~\cite{desroches2013bursting, vo2013pseudoplateau}. For a comprehensive overview of canards, we refer to the books~\cite{dumortier1996cycles,de2021canard}. For our study, the most relevant articles related to canards are~\cite{benoit2002combined, callot1985visibles, callot1993champs}, the PhD thesis by J.-L. Callot~\cite{callot1981bifurcations} and the articles on canards in 2D torus systems~\cite{guckenheimer2001duckdevil, schurov2010torus, schurov2017factory}

Among a large number of references related to the exact WKB method and its applications, we highlight the following reader-friendly papers~\cite{sueishi2020quantization, sueishi2021anomaly}, where the basic ideas of non-perturbative effects are presented, and papers~\cite{dunne2017mathieu, basar2017quantumgeometry}, where the examples of computations are provided. We especially highlight papers~\cite{nikolaev2023riccati, nikolaev2022existence, nikolaev2023exactwkb, nikolaev2024geometry, nikolaev2024gevrey}, where the rigorous study of singular perturbation theory for Riccati and Schr\"{o}dinger equations was done. We also want to draw attention to recent papers in which connections between slow-fast systems and WKB theory were presented for several examples~\cite{kristiansen2024wkbturning, kristiansen2025wkbeigenvalue, kristiansen2026codim2}. The singular perturbation theory goes hand in hand with asymptotic series analysis, resurgence and Stokes phenomena, Borel resummation, etc. Among physicists, these effects are typically associated with instantons, while in mathematics one prefers to speak about \emph{alien calculus}~\cite{dorigoni2019introduction} and trans-series~\cite{aniceto2019primer, ecalle1993transseries}.

The main passage of our paper is two following statements,
\begin{equation*}
\begin{gathered}
    \text{canards}=\text{non-perturbative eff. in asympt. series} \\
    \text{log of canard window}=\text{scaled instanton action}
\end{gathered}
\end{equation*}
Our main goal is to provide comprehensive narration, avoiding the overload of formal and technical details (it is unavoidable, but we have tried to push all technical details into the Appendix). Our key result is the catalog of canards of different types, whose origin is almost the same: an accurately computed instanton action. Having described the general picture, let us provide the essential facts about the canards and dynamical systems on the 2D torus. 

\subsection{Slow-fast systems and canards}

The name ``canard'' is just a loanword French ``duck'' (it is funny to notice that ``canard'' also means a hoax in French). The canard trajectory describes certain behavior of the trajectory in the slow-fast dynamical system. One of the simplest possible slow-fast systems is the Van der Pol (VdP) oscillator (here we follow~\cite{desroches2011curvature}),
\begin{equation}
    \epsilon\ddot{x}+\left(x^2-1\right)\dot{x}+x-a=0,
\end{equation}
where $0<\epsilon\ll 1$ and $a$ is the additional parameter. It is a Li\'{e}nard system, which can be written as the slow-fast system with variable $y=\epsilon\dot{x}+(x^3/3-x)$ and time rescaling $\tau=t/\epsilon$,
\begin{equation}
\begin{cases}
    x'=y-x^3/3+x,\\ 
    y'=\epsilon(a-x).
\end{cases}
\end{equation}
Here $y$ is slow, $x$ is fast, $x'=dx/d\tau$. In the singular limit $\epsilon=0$, the slow (or critical) curve is given by
\begin{equation}
    \mathcal{C}_0=\left\{(x,y):y-\frac{x^3}{3}+x=0\right\}
\end{equation}
and consists of two stable (attracting) and one unstable (repelling) branches, $\mathcal{C}_0 = \mathcal{C}^{a}_0\cup\mathcal{C}^{r}_0$. The attracting branch means $\partial_x f<0$, while the repelling means $\partial_x f>0$ with $f(x,y)=y-x^3/3+x$. The Fenichel theory guaranties that away from the fold points, where $\partial_xf=0$, the perturbed by $\epsilon\neq 0$ branches $C_{\epsilon}^{a}$ and $C_{\epsilon}^{r}$ are $O(\epsilon)$-close to $\mathcal{C}_0^{a}$ and $\mathcal{C}_0^{r}$. These are locally invariant slow manifolds, which are not unique, but any two choices are $O(e^{-c/\epsilon})$-close, so one speaks about the slow manifold up to exponentially small ambiguity. By definition, a canard trajectory is a trajectory that \emph{rides the unstable branch} for a slow-time interval of $O(1)$-order. 

\begin{figure}
    \centering
    \includegraphics[width=\linewidth]{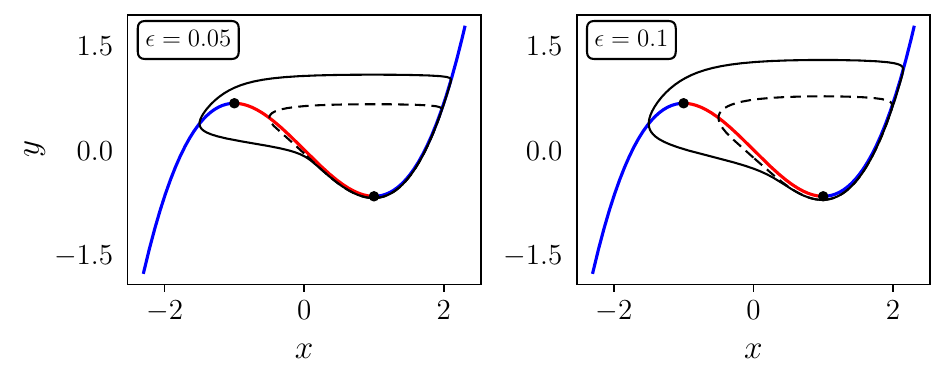}
    \caption{Canard trajectories in VdP system. Solid line: canard with ``head'', dashed line: canard without ``head''. Unstable branch of critical curve is colored red, stable branch is colored blue. Fold points are shown as black dots.}
    \label{fig:vdp-canards}
\end{figure}

To capture a duck, i.e., to find a canard trajectory, we need to have appropriate hunting equipment. The hunt was opened in~\cite{benoit1981chasse} using nonstandard analysis, and in~\cite{eckhaus1983relaxation} by matched asymptotics; the geometric formulation was developed in~\cite{krupa2001relaxation}. The analytic idea is to use the definition of maximal canard, which is an orbit $\mathcal{C}_{\epsilon}^{a}\cap\mathcal{C}_{\epsilon}^{r}$. We continue $\mathcal{C}_{\epsilon}^{a}$ forward and $\mathcal{C}_{\epsilon}^{r}$ backward by the flow into the fold region, and then measure the signed distance $d$ on a transverse section. This distance is a function of $a$ and $\sqrt{\epsilon}$, i.e. $d=d(a,\sqrt{\epsilon})$. The blow-up analysis at the canard point~\cite{krupa2001fold} shows that $d$ extends to the function of $a$ and $\sqrt{\epsilon}$ that is smooth up to $\epsilon=0$ and regular in $a$. Hence, the maximal canard exists if and only if $d(a,\sqrt{\epsilon})=0$, which defines a unique curve. The blow-up method underlying this construction originates in~\cite{dumortier1996cycles}, also see~\cite{jardonkojakhmetov2021blowup} for a survey.

\subsection{Systems on 2D torus}

In this paper, we focus on the dynamical systems on 2D torus. The reason for this is, in a sense, technical: it is known (shown and proven) that for such systems the canard trajectories exist without any additional parameters~\cite{guckenheimer2001duckdevil}. Moreover, we consider the special class of these systems. The dynamical system on 2D torus can be represented as
\begin{equation}
\begin{cases}
    \dot{\phi}=F(\tau,\phi),\\
    \dot{\tau}=\omega.
\end{cases}
\end{equation}
Here, $F(\tau,\phi)$ is the $2\pi$-periodic function in each variable. The phase space is $\mathbb{T}^2=S^1_{\phi}\times S^1_{\tau}$. We consider the class of systems where the Fourier expansion of $F(\tau,\phi)$ with respect to $\phi$ contains only the first modes. We call such systems \emph{M\"{o}bius systems} (as we will see further, this property plays a crucial role). Each system on 2D torus has the nice topological invariant, the Poincar\'{e} rotation number,
\begin{equation}
    \rho=\lim\limits_{t\to\infty}\frac{\phi(t)-\phi(0)}{\omega t},
\end{equation}
where $\phi(t)$ represents an unwrapped trajectory (i.e., we consider the lift of $\phi$ to the real line). The type of a trajectory on the torus is characterized by $\rho$. The rotation number can also be computed from the first recurrence map, $\mathcal{P}:S^1_{\phi}\to S^1_{\phi}$ as a limit of consecutive action of $\mathcal{P}$ on $\phi(0)\in S^1_{\phi}$. For the M\"{o}bius systems, the first recurrence map is M\"{o}bius transformation that maps $S^1$ to itself. Such M\"{o}bius transformation can have a stable fixed point on $S^1$ (this means phase-locking and $\rho$ is integer) or do not have fixed points on $S^1$ ($\mathcal{P}$ is rotation, no phase-locking). Since for a general system, phase-locking typically corresponds to rational $\rho$, the integrality of $\rho$ is called \emph{rotation number quantization effect}~\cite{buchstaber2010rotation} (and it is true quantum-mechanical quantization, which was shown in~\cite{alexandrov2026phaselocking}). Domains in parameter space where the phase-locking occurs are called Arnold tongues. If the reader wants to understand better the mentioned facts, all the relevant details are clarified in the Appendix~\ref{app:torus}.

 Our workhorse is the following chain of equivalences,
\begin{equation*}
    \text{M\"{o}bius system}\leftrightarrow\text{Riccati}\leftrightarrow\text{Schr\"{o}dinger}
\end{equation*}
Separately, each element of this duality chain has its own caveats, which complicate complete understanding. Switching between these three pictures allows flexibility and simplifies technical moments. The structure of the paper is following. We start with a simple but physically relevant example of the M\"{o}bius system on the torus that represents the overdamped limit of the Josephson junction. We provide a detailed analysis of trajectory behavior in the gaps between Arnold tongues, then obtain the complete description of Arnold tongues in terms of quantum mechanics. We utilize our quantum-mechanical picture to locate the region of canard existence and to investigate canard trajectories. We finalize the main part of paper by providing the purely classical point of view on the asymptotic series in slow-fast systems theory and their non-perturbative effects. We complete the paper with a discussion of possible connections to well-known phenomena in Hamiltonian mechanics and mention further possible research directions.

\section{Josephson junction}

The resistively shunted Josephson junction (JJ) in the overdamped limit can be described in terms of the RSJ model~\cite{Likharev1986,BaronePaterno1982},
\begin{equation}\label{eq:RSJ-model}
    \dot{\phi}=-\sin\phi+B+A\cos\omega t.
\end{equation}
Here, all quantities are dimensionless; $A$ and $B$ correspond to the AC and DC amplitudes of the external current, respectively; $\omega$ represents the frequency of the AC current. It is well-known that Poincar\'{e} rotation number for this system is up to constant $\hbar\omega/(2e)$ coincides with the average voltage on JJ. To be precise,
\begin{equation*}
    \langle\mathcal{V}\rangle = \frac{\hbar}{2e}\langle\dot{\phi}\rangle,\quad \langle\dot{\phi}\rangle = \omega\rho,
\end{equation*}
where $\langle\cdot\rangle$ means time-average and $\mathcal{V}$ denotes the voltage. This relation is nothing more than Shapiro steps~\cite{Shapiro1963}. The main reason for making this choice is simple: JJs are intensively studied in experiments and look prominent in practical applications. The voltage-current curve is one of the simplest quantities to measure, for example, in papers~\cite{panghotra2020giant,stolyarov2026shapiro} the Arnold tongues for RSJ model are obtained experimentally. It is important to emphasize that this model~\eqref{eq:RSJ-model} has already appeared in the context of adiabatic limit $\omega\to 0$ and our study extends the results in~\cite{gandhi2015dynamics} and provides a more general framework.

\begin{figure}
    \centering
    \includegraphics[width=0.9\linewidth]{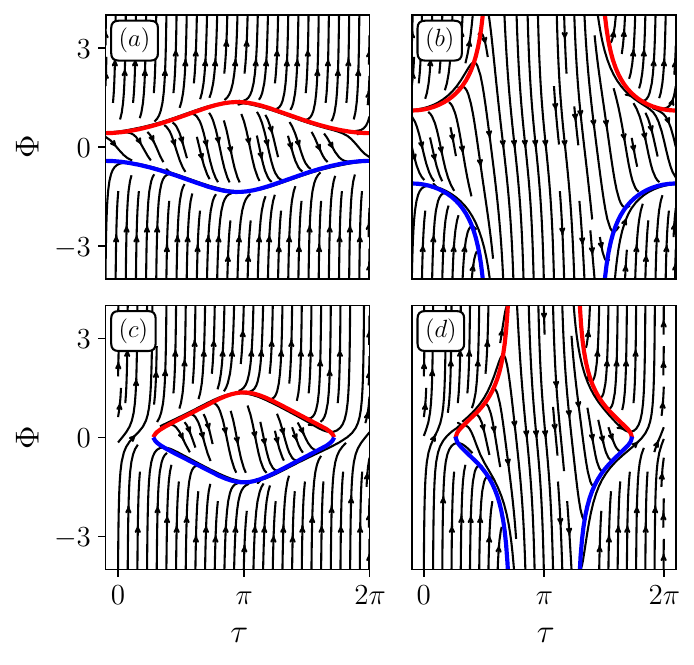}
    \caption{Vector fields for eq.~\eqref{eq:RSJ-Ricc} with $\omega=0.1$. Blue line corresponds to stable branch $\Phi_0^{-}(\tau)$, red line corresponds to unstable branch $\Phi_0^{+}(\tau)$. Parameters are: (a) $A=0.5$, $B=0.2$, (b) $A=0.7$, $B=-0.8$, (c) $A=0.8$, $B=0.5$, (d) $A=1.5$, $B=0.0$. Cases (b-d) are ``canard candidates'': at least one real fold point exists. For (a) there is no real fold points. In all cases the branches continue by periodicity in $\tau$.}
    \label{fig:riccati_flow}
\end{figure}

Introducing the variable $\tau=\omega t$, we have the dynamical system on a 2D torus, $\mathbb{T}^2=S^1_{\phi}\times S^1_{\tau}$. It is convenient to shift $\phi\to\phi+\pi/2$ and then rewrite~\eqref{eq:RSJ-model} in terms of the variable $\Phi=\tan(\phi/2)$,
\begin{equation}\label{eq:RSJ-Ricc}
    \omega\frac{d\Phi}{d\tau}=f_{+}(\tau)\Phi^2-f_{-}(\tau),
\end{equation}
where we have introduced
\begin{equation*}
    f_{\pm}(\tau)=\frac{1}{2}\pm\frac{B+A\cos\tau}{2}.
\end{equation*}
In the limit $\omega\to 0$, the variable $\tau$ is slow, $\Phi$ is fast (or $\phi$, if we work in terms of~\eqref{eq:RSJ-model}). The critical curve is defined by the equation $\dot{\Phi}=0$, which gives
\begin{equation}\label{eq:critical-curve}
    \Phi(\tau)^2=\frac{f_{-}(\tau)}{f_{+}(\tau)}.
\end{equation}
It contains two branches: stable $\Phi_0^{-}(\tau)$ and unstable $\Phi_{0}^{+}(\tau)$,
\begin{equation*}
    \Phi^{-}_0(\tau)=-\sqrt{\frac{f_{-}(\tau)}{f_{+}(\tau)}},\quad\Phi^{+}_0(\tau)=+\sqrt{\frac{f_{-}(\tau)}{f_{+}(\tau)}}.
\end{equation*}
For $\omega=0$, there are two special points $\tau_{\pm}$, which are zeros of the functions $f_{\pm}(\tau)$. At the point $\tau_{-}$, two branches $\Phi_0^{\pm}$ coincide (see fig.~\ref{fig:riccati_flow}c,~\ref{fig:riccati_flow}d). At the point $\tau_{+}$, we naively see the blow-up, but change $\Phi\to 1/\Phi$ clearly shows that again two branches coincide. The flow of vector field is shown at fig.~\ref{fig:riccati_flow}
It is easy to see that the real fold point does not exist for all the values of $B$ and $A$. For $0<\omega\ll 1$ the branches perturb invariant curves $\Phi^{\pm}(\omega,\tau)$, obtained by Fenichel's theorem on any arc where $f_{+}(\tau)f_{-}(\tau)\neq 0$. Each branch admits a unique formal expansion,
\begin{equation}
    \Phi^\pm(\omega,\tau)=\sum_{k\ge0}\Phi^\pm_k(\tau)\,\omega^k,\quad \Phi_0^{\pm}=\pm\sqrt{\frac{f_{-}(\tau)}{f_{+}(\tau)}}
\end{equation}
with $\Phi_0^{\pm}(\tau)=\pm\sqrt{f_{-}(\tau)/f_{+}(\tau)}$. The coefficients $\Phi_{k}(\tau)$ acquire poles of growing order at the folds and this expansion is valid only away from $\tau_{\pm}$. Moreover, it is divergent, Gevrey-1 series~\cite{canalisdurand2000gevrey}, whose divergence is governed by the complex zeros of $f_{\pm}(\tau)$ (similar to complex turning points in~\cite{benoit1998surstables}). Therefore, $\Phi^{-}$ and $\Phi^{+}$ are determined only up to $O(e^{-c/\omega})$. This means that the forward continuation of $\Phi^{-}$ past a fold point differs from the backward continuation of $\Phi^{+}$ by an exponentially small quantity~\cite{fruchard2013composite} (and~\cite{fruchard2009overstability} for review). Canard trajectories arise for exactly such parameters values at which this splitting vanishes.

Knowing these facts, we want to derive a criterion for the existence of canard solutions. To start with, we need to develop a framework that will allow us to compute all the quantities needed. The framework will be based on elliptic curves, and differential form on these curves.

\subsection{Critical curve is elliptic curve}

Let us consider the expression for the critical curve,
\begin{equation}\label{eq:crit-ell-curve}
    \Phi^2(\tau)=\frac{1-B-A\cos\tau}{1+B+A\cos\tau}.
\end{equation}
It is easy to see that $\Phi_0^{-}(\tau)$ and $\Phi_0^{+}(\tau)$ are in fact two branches of the same function in the complex plane. Setting $u=\tan(\tau/2)$, one can see that~\eqref{eq:crit-ell-curve} defines an elliptic curve, which is topologically equivalent to the torus. On the torus, we have two homology cycles. The first cycle, $\alpha$-cycle, encircles points $-\tau_{-}$ and $\tau_{-}$, where $\tau_{-}$ is zero of $f_{-}(\tau)$. The second cycle, $\beta$-cycle, encircles points $\tau_{-}$ and $\tau_{+}$, where $\tau_{+}$ is zero of $f_{+}(\tau)$. Throughout the whole text, the cycle integrals are taken on the double cover, i.e.
\begin{equation*}
    \oint_{\alpha}=2\int_{\text{$\alpha$-cut}},\quad \oint_{\beta}=2\int_{\text{$\beta$-cut}}.
\end{equation*}
As should be, this elliptic curve has the holomorphic 1-form, which tells us about the geometry of our torus. But the main object of our interest is the following 1-form,
\begin{equation}
    \Omega=2\sqrt{f_{-}(\tau)f_{+}(\tau)}\,d\tau,
\end{equation}
which, as we will see below, has a nice dynamical interpretation. In fact, we deal with two slightly different forms which are proportional to each other but have different physical interpretations. The detailed derivation of the fact that~\eqref{eq:crit-ell-curve} is indeed an elliptic curve with computation of torus periods and other related objects is given in the Appendix~\ref{app:ell-curve}.

\subsection{Fold points under scrutiny}

Fold points play a significant role in our framework: they carry the Maslov index. We need to know what happens to a trajectory that follows the stable branch of the slow curve, reaches a fold, and is released into the interval where the slow curve has no real branches.

Without loss of generality, let $\tau_{+}$ be complex, so that the only real folds are $\pm\tau_{-}$: the trajectory leaves the slow curve at $\tau_{0}=-\tau_{-}$ and is recaptured at $+\tau_{-}$. Near $\tau_{0}$ we can expand $f_{-}(\tau)\approx a(\tau_{0}-\tau)$ with $a=-f_{-}'(\tau_{0})>0$, and $f_{+}(\tau_{0})=1$, the latter because $f_{+}(\tau)+f_{-}(\tau)=1$. Introducing the rescaled variables
\begin{equation}\label{eq:fold-scaling}
    \tau-\tau_{0}=\omega^{2/3}\xi,\qquad \Phi=\omega^{1/3}\eta,
\end{equation}
which place the slow region at $\xi<0$, brings \eqref{eq:RSJ-Ricc} to the local
normal form
\begin{equation}\label{eq:fold-normal-form}
    \frac{d\eta}{d\xi}=\eta^{2}+a\xi .
\end{equation}
The slow curve is $\eta=\pm\sqrt{a|\xi|}$, $\xi\le 0$. Physical time runs towards increasing $\xi$, so the branch attracting in $\tau$ is $\eta_{-}=-\sqrt{a|\xi|}$, i.e.\ $\Phi=-\sqrt{f_{-}/f_{+}}$. A trajectory that reaches the fold along it is the unique solution of \eqref{eq:fold-normal-form} with $\eta\to-\sqrt{a|\xi|}$ as $\xi\to-\infty$; in physical time, all nearby solutions collapse onto it exponentially fast. It is
\begin{equation}\label{eq:eta0}
    \eta_{0}(\xi)=a^{1/3}\,
    \frac{\mathrm{Ai}'(-a^{1/3}\xi)}{\mathrm{Ai}(-a^{1/3}\xi)},
\end{equation}
and all we need is its behavior on the far side of the fold,
\begin{equation}\label{eq:eta-asym}
    \eta_{0}(\xi)\sim\sqrt{a\xi}\,
    \tan\left(\frac{2\sqrt{a}}{3}\xi^{3/2}-\frac{\pi}{4}\right),\qquad
    \xi\to+\infty .
\end{equation}
Returning to the original variables, we can write
\begin{equation}\label{eq:after-fold}
\begin{gathered}
    \Phi(\tau)\sim\sqrt{-\frac{f_{-}(\tau)}{f_{+}(\tau)}}
    \tan\left(\theta-\frac{\pi}{4}\right),\\
    \theta=\frac{1}{\omega}\int_{\tau_0}^{\tau}d\tau'
    \sqrt{-f_{+}(\tau')f_{-}(\tau')}
\end{gathered}
\end{equation}
where $\tau>\tau_0$ and the amplitude reduces to $\sqrt{a(\tau-\tau_{0})}$ close to the fold. Had we ignored the fold and counted the phase from $\tau_{0}$, we would have obtained $\tan\theta$. The trajectory is therefore \emph{delayed} by $\pi/4$: it first crosses $\Phi=0$ at $\theta=\pi/4$ rather than at $\theta=0$. The mirror fold $+\tau_{-}$ is treated identically, since $(\tau,\Phi)\mapsto(-\tau,-\Phi)$ is a symmetry of \eqref{eq:RSJ-Ricc} (the coefficients $f_{\pm}$ are even) that interchanges one fold with another; in the rescaled variables it acts as $(\xi,\eta)\mapsto(-\xi,-\eta)$ and carries \eqref{eq:fold-normal-form} into $\eta'=\eta^{2}-a\xi$, with the slow region now at $\xi\ge 0$. The Appendix~\ref{app:maslov} analyzes \eqref{eq:fold-normal-form} itself, which therefore covers both folds.

\begin{figure}
    \centering
    \includegraphics[width=0.75\linewidth]{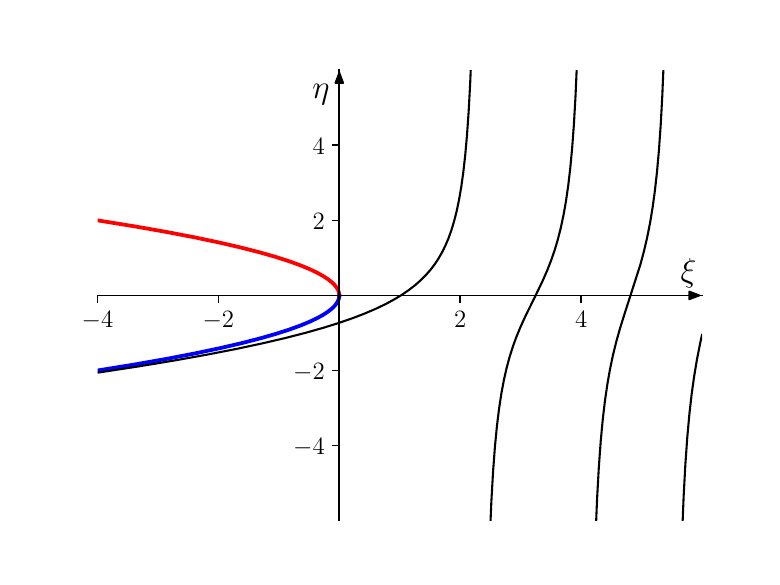}
    \caption{Local dynamics near the fold. The attracting branch is colored blue, the repelling branch is colored red.}
    \label{fig:airy-traj}
\end{figure}

Therefore, each simple fold contributes $\pi/4$ and a closed trajectory acquires
$2\times\pi/4=\pi/2$, the Maslov index $1/2$. For folds of the second type,
$\pm\tau_{+}$ with $f_{+}(\pm\tau_{+})=0$, which are the real folds in the complementary case, the analysis is unchanged after $\Phi\to 1/\Phi$, which swaps $f_{-}$ and $f_{+}$.

It is worth being explicit about what selected \eqref{eq:eta0}: only recessiveness at the fold, i.e. the demand that $\eta$ approach a definite branch as $\xi\to-\infty$. Nothing in the derivation requires the trajectory to track an unstable branch for a $O(1)$ time. At $\tau_{0}$, where the trajectory
is released, the branch along which it arrives is the attracting one, so $\eta_{0}$ is simply the ordinary relaxation trajectory; an arbitrary solution differs from it by $O(e^{-2S/\omega})$, with $S=\int\sqrt{f_{+}f_{-}}\,d\tau$ over the preceding slow interval, which does not affect the phase in the leading order. At the mirror fold $+\tau_{-}$, where the trajectory is captured, the symmetry $(\tau,\Phi)\mapsto(-\tau,-\Phi)$ exchanges the two branches and the same $\eta_{0}$ describes a canard, i.e. the trajectory must be recessive at both ends. Thus, $\pi/4$ is a property of the fold and of the recessive solution there, identical in the two cases, which allows it to be attached to each fold separately. It is the well-known consequence of the Stokes phenomenon and can be obtained from the WKB connection formulas, which was also presented the Appendix~\ref{app:maslov}.

\subsection{Node counting procedure}

What happens when we leave the stable branch through the fold point? In fact, nothing complicated: $\Phi$ evolved according to the equations of motion. Let us consider the most general case, when both fold points $\tau_{-}$ and $\tau_{+}$ are real. In this case, we have the following ordering,
\begin{equation*}
    0<\tau_{-}<\tau_{+}<\pi<2\pi-\tau_{+}<2\pi-\tau_{-}<2\pi.
\end{equation*}
In domains $[0,\tau_{-}]$ and $[2\pi-\tau_{-},2\pi]$ we have $f_{-}(\tau)<0$. In domain $[\tau_{+},2\pi-\tau_{+}]$ we have $f_{+}(\tau)<0$. Hereafter, we will use the following names:
\begin{equation*}
\begin{gathered}
    \text{lens}=\text{region where slow-manifold is real},\\
    \text{well}=\text{region where $\Phi$ runs},\\
\end{gathered}
\end{equation*} 
In these domains, away from the stable branch, our $\Phi$ rides from $-\infty$ to $+\infty$ and crosses zero. We want to understand how many nodes, i.e. zero crossings, we can have. In the domain $[0,\tau_{-}]$, $\Phi$ runs to $+\infty$ and returns from $-\infty$. The slow-time required for this event is
\begin{equation*}
    \omega\int_{-\infty}^{+\infty}\frac{d\Phi}{f_{+}(\tau)\Phi^2+|f_{-}(\tau)|}=\frac{\omega\pi}{\sqrt{-f_{-}(\tau)f_{+}(\tau)}}.
\end{equation*}
In the domain $[\tau_{+},\pi]$, $\Phi$ runs to $-\infty$ and returns from $+\infty$, so again the required slow-time is
\begin{equation*}
    \omega\int_{-\infty}^{+\infty}\frac{d\Phi}{|f_{+}(\tau)|\Phi^2+f_{-}(\tau)}=\frac{\omega\pi}{\sqrt{-f_{-}(\tau)f_{+}(\tau)}}.
\end{equation*}
Notice that these formulas are valid in the adiabatic limit, i.e. everywhere except neighborhoods of fold points. Also, by symmetry, we can consider $[0,\tau_{-}]$ and $[\tau_{+},\pi]$ intervals, then just double the number of nodes.

In terms of the original phase variable $\phi$, these events are simply \emph{phase-slips} and in fact we have estimated phase-slip rates. Notice that each phase-slip event requires the same slow-time, but in domains $[0,\tau_{-}]$ and $[\tau_{+},\pi]$ we have opposite directions of phase-slips. The accumulated phases for $[-\tau_{-},\tau_{-}]$ (this corresponds to the union of $[0,\tau_{-}]\cup[2\pi-\tau_{-},2\pi]$) and $[\tau_{+},2\pi-\tau_{+}]$ are given by
\begin{equation}
\begin{gathered}
    N_{>}(A,B)=\frac{2}{\omega}\int_{0}^{\tau_{-}}d\tau\sqrt{-f_{-}(\tau)f_{+}(\tau)},\\
    N_{<}(A,B)=\frac{2}{\omega}\int_{\tau_{+}}^{\pi}d\tau\sqrt{-f_{-}(\tau)f_{+}(\tau)},
\end{gathered}
\end{equation}
where we emphasize that these quantities depend on $A$ and $B$. Recall that each domain $[-\tau_{-},\tau_{-}]$ and $[\tau_{+},2\pi-\tau_{+}]$ is flanked by two fold points, and each fold point delays the phase by $\pi/4$ phase-slip (Maslov index). Then, if the number of nodes is an integer in $[-\tau_{-},\tau_{-}]$ or in $[\tau_{+},2\pi-\tau_{+}]$, we will have
\begin{equation}
    N_{>}=\pi\left(n+\frac{1}{2}\right)\text{ or }N_{<}=\pi\left(m+\frac{1}{2}\right),\quad n,m\in\mathbb{Z}_{+}.
\end{equation} 
Both fold points are real if $|A|\geq 1+|B|$. On the interval $[-\tau_{-},\tau_{-}]$ we have $B+A\cos\tau>1$, while on the interval $[\tau_{+},2\pi-\tau_{+}]$ we have $B+A\cos\tau<-1$. Let $q=B+A\cos\tau$, then the branch of $\sqrt{q^2-1}$ behaves as $q$ as $|q|\to\infty$ and is analytic and periodic in the upper half-strip $\im \tau>0$. Deforming upper half-strip $\tau$-contour to $\im\tau\to\infty$, we obtain the following identity (it is \emph{exact} and holds for $|A|>|B|+1$),
\begin{equation}
    \omega\left(N_{>}-N_{<}\right)=\pi B.
\end{equation}
Note that if we want to have \emph{both} $N_{>}$ and $N_{<}$ quantized, we have to force,
\begin{equation}
    \omega(n-m)=B \Leftrightarrow B=l\omega,l\in\mathbb{Z}
\end{equation}
and this result was proven with tremendous technicalities~\cite{glutsyuk2014adjacency, glutsyuk2019constrictions}. The mystery about the so-called ``constrictions'' (points in $(B,A)$-space, where the Arnold tongue shrinks to a point) disappears: constriction means double resonance. In terms of quantum mechanics, it is a simple thing: we have a diatomic lattice (two wells), so we have two sublattices, their bands would repel, and a minigap will appear. This effect is protected by the exact residue relation, which receives no $\omega$-corrections.

The proposed analysis is needed to connect the obtained quantities $N_{>}$ and $N_{<}$ to the integrals over the homology cycles of the elliptic curve. It was our preliminary reconnaissance for the ducks: we have seen that the canard trajectory inevitably has $\pi/2$ phase shift, since it is associated with two fold points.

\subsection{Balancing trajectories}

Now consider the local contraction rate $\mu(\tau)$ of ``fiber'' in the phase space, which is given by
\begin{equation}\label{eq:exact-Floquet}
    \mu(\tau,\omega)=\frac{\partial F(\Phi,\tau)}{\partial\Phi}=2f_{+}(\tau)\Phi(\omega,\tau),
\end{equation}
where $\Phi(\omega,\tau)$ is the exact slow manifold and $F(\Phi,\tau)=f_{+}(\tau)\Phi^2-f_{-}(\tau)$. The ``exact'' means that we have to understand eq.~\eqref{eq:exact-Floquet} as the series with respect to $\omega$. It is not difficult to notice that~\eqref{eq:exact-Floquet} is nothing more than the \emph{instantaneous} Floquet exponent. The usual Floquet exponent is obtained after the integration of $\mu(\tau)$ over the period. In the leading order in $\omega$, we have simply
\begin{equation*}
    \mu(\tau,\omega)=\mu_0(\tau)=2f_{+}(\tau)\Phi_0(\tau)=\pm 2\sqrt{f_{-}(\tau)f_{+}(\tau)}.
\end{equation*}
Two different signs reflect that the monodromy belongs to $\mathrm{SL}(2,\mathbb{R})$ and has reciprocal multipliers

\begin{figure}
    \centering
    \includegraphics[width=\linewidth]{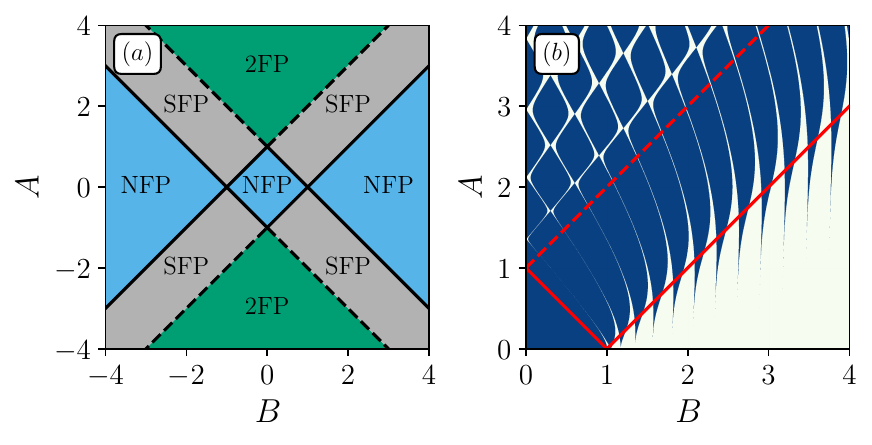}
    \caption{(a) Regimes chart: crossing dashed black lines causes disappear of real $\tau_{+}$, $\tau_{-}$ remains real; crossing solid black lines causes disappear of real $\tau_{-}$; in 2FP (filled by green) domain we have both $\tau_{\pm}$ real; in SFP (filled by grey) only one of $\tau_{\pm}$ is real; in NFP (filled by blue) no real fold points; (b) Arnold tongues structure with $\omega=0.3$ in positive $(B,A)$-quarter (tongues are dark): above dashed red line double resonance is possible; between dashed and solid red lines only one well is quantized; below solid red line gaps becomes very thin.}
    \label{fig:regimes-chart}
\end{figure}

In order to have balanced trajectory, the accumulated real part of the Floquet exponent in oscillatory domains, i.e., precisely in the $\alpha$-cycle, should be zero. Obviously, due to the oscillatory nature, the imaginary part is not zero. These statements can be formalized as
\begin{equation}
\begin{gathered}
    \re\oint_{\alpha}d\tau\,\mu(\tau,\omega)=0,\\
    \frac{1}{2}\im\oint_{\alpha}d\tau\,\mu(\tau,\omega)=2\pi\omega\left(n+\frac{1}{2}\right),
\end{gathered}
\end{equation}
where $n\in\mathbb{Z}_{+}$. The first is tautological. Here we just recall our $\alpha$-cycle definition and recognize the proposed 1-form $\Omega$. The last expression just tells us that \emph{one of} the quantities $N_{>}$ or $N_{<}$ is quantized, but does not tell \emph{which one}. The constant shift $\pi B/\omega$ discussed earlier is nothing more than the pole of our form. These conditions correctly reflect the fact that $\mu$ is the same function on different branches. Notice that these expressions are \emph{exact}, since $\mu$ is defined via the exact slow-manifold, and we just omit the $\omega$-dependence for brevity. In the leading order in $\omega$, we have
\begin{equation*}
    \mu_0(\tau)=\sqrt{1-(B+A\cos\tau)^2}.
\end{equation*}
and the leading-order quantization is
\begin{equation}
    \int_{-\tau_{-}}^{\tau_{-}}d\tau\,\sqrt{(B+A\cos\tau)^2-1}=2\pi\omega\left(n+\frac{1}{2}\right)
\end{equation}
with $n\in\mathbb{Z}_{+}$. The computation of this integral depends on the relation between $B$ and $A$. The complete regime description is given in tab.~\ref{tab:FP-regimes} and shown in fig.~\ref{fig:regimes-chart}(a).
\begin{table}[h!]
\begin{tabular}{c|c|c}
regime & fold points                                      & condition               \\ \hline
2FP    & $\tau_{\pm}\in\mathbb{R}$                        & $|A|-|B|\geq 1$ \\ \hline
SFP    & $\tau_{-}\in\mathbb{R}$, $\tau_{+}\in\mathbb{C}$ & $|1-B|\leq |A|\leq 1+B$, $B>0$ \\ \hline
SFP    & $\tau_{+}\in\mathbb{R}$, $\tau_{-}\in\,\mathbb{C}$ & $|1+B|\leq|A|\leq 1-B$, $B<0$  \\ \hline
NFP    & $\tau_{\pm}\in\mathbb{C}$                        & $|A|\leq||B|-1|$                   \\
\end{tabular}
\caption{Fold points regimes}
\label{tab:FP-regimes}
\end{table}

\begin{figure}
    \centering
    \includegraphics[width=0.75\linewidth]{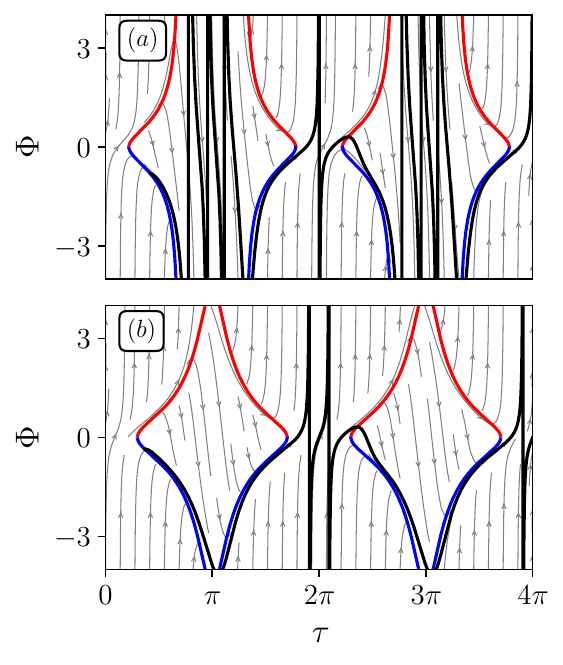}
    \caption{(a) $\Phi(t)$ for $\omega=0.1$, $n=1$, $A=1.5$ (respective $B=-0.15894$), both $\tau_{\pm}$ are real; (b) $\Phi(t)$ for $\omega=0.1$, $n=2$, $A=1.2$ (respective $B=0.29034$), $\tau_{+}$ is complex, $\tau_{-}$ is real and causes quantization.}
    \label{fig:quantized-trajs}
\end{figure}

Notice that a very nice and specific symmetry exists: if we change $B\to-B$, then we effectively replace $\tau\to \pi-\tau$ and change $\tau_{-}\to\tau_{+}$. This means that we can evaluate the integral
\begin{equation}\label{eq:leading-order-int}
    I=\int_{0}^{\tau_{-}}d\tau\sqrt{(B+A\cos\tau)^2-1},
\end{equation}
and then use the change $B\to-B$ to obtain the complete picture. The integral~\eqref{eq:leading-order-int} can be computed explicitly and expressed via the elliptic functions (see the Appendix) and in the leading order in $\omega$ we can write
\begin{equation*}
    I=\pi\omega\left(n+\frac{1}{2}\right),\quad n\in\mathbb{Z}_{+}.
\end{equation*}
We have matched our earlier observations on node counting with the local contraction rate $\mu(\tau)$. Now, let us visualize our formulas. We set a small value of $\omega$, then use the explicit value of $I(A,B)$ and the quantization condition to obtain the set of curves in the $(B,A)$-plane for different values of $n$. These curves allow us to find pairs $(B,A)$ that obey the quantization condition and then use these values to find trajectories $\Phi=\Phi(t)$, see fig.~\ref{fig:quantized-trajs}.

The whole $(B,A)$-plane consists of the following domains: domain, where both $\tau_{-}$ and $\tau_{+}$ are real (two fold points domain, 2FP); only $\tau_{-}$ is real (single fold point, SFP); no real fold points (NFP). In 2FP domain, double-resonance condition can be satisfied, i.e. in this domain there exist ``constrictions''. In the SFP domain, only $\tau_{-}$ (or $\tau_{+})$ contributes to the quantization condition. This picture is perfectly symmetric: by swapping $B\leftrightarrow-B$ (it gives $\tau\to\pi-\tau$ change), we effectively swap $\tau_{-}\leftrightarrow\tau_{+}$; by swapping $A\leftrightarrow-A$, we again swap $\tau_{\pm}\leftrightarrow\pi-\tau_{\pm}$. All these relations simply represents the symmetries $(B,A)\leftrightarrow(-B,A)$ and $(B,A)\leftrightarrow(B,-A)$.

We have discussed in detail what happens in the 2FP and SFP domains but have not discussed the NFP domain. From the fig.~\ref{fig:regimes-chart}, we see that NFP contains two subdomains: NFP-c (central ``diamond'', where $|A|+|B|<1$) and NFP-t (two ``tails'', where $|A|<|B|-1$). In the classical picture, we do not see any significant difference between these two subdomains, but their behaviors are very different in quantum picture. Here, both $\tau_{\pm}$ are complex and we have two subregions: central ``diamond'' and two ``tails''. If $|B|>1$, we are in tails and here our fold points are
\begin{equation*}
\begin{gathered}
    \tau_{\pm}=\pi+i\arccosh\left|\frac{B\pm1}{A}\right|,\,\sgn(AB)=1,\\
    \tau_{\pm}=-i\arccosh\left|\frac{B\pm 1}{A}\right|,\,\sgn(AB)=-1.
\end{gathered}
\end{equation*}
If $|B|<1$, we are in the central diamond, and here
\begin{equation*}
    \tau_{\pm}=\frac{\pi(1\pm\delta)}{2}\mp i\delta\arccosh\left(\frac{1\mp B}{A}\right),\,\,\delta=\sgn(A).
\end{equation*}
These expressions tell us that the $\alpha$-cycle integral obtains the imaginary contribution when we are in the NFP regime. The role of this imaginary contribution will be clear in our quantum mechanical description. In terms of our torus, crossing the line $|A|=|B|-1$ means that the roles of $\alpha$- and $\beta$-cycles interchange.

\begin{figure}
    \centering
    \includegraphics[width=0.9\linewidth]{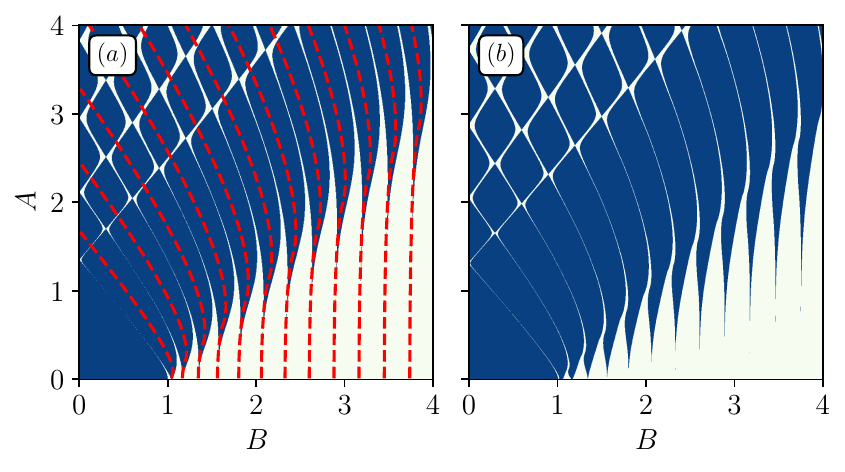}
    \caption{(a) Numerically obtained Arnold tongues (shaded by dark) and its center location from leading order analytical expressions (dashed red), (b) analytical leading order band-gap structure, obtained from DDP quantization. Both pictures have $\omega=0.3$.}
    \label{fig:bandgap-leading-order}
\end{figure}

The last point that we need to say to be honest. First, our quantization condition, obtained from the node counting procedure, determines \emph{centers} of domains, where the phase-locking is absent. We can also locate the centers of the Arnold tongues, but we postpone the explanation why it works for further narration, see fig.~\ref{fig:bandgap-leading-order}. Second, the presented trajectories touch the unstable branch of the slow-curve, but of course they are not canard trajectories. They just represent relaxational oscillations.

The developed description and observations give us the hint: to capture canards, we need to use elliptic curve language and carefully deal with differential forms on it. And, of course, we know the missing ingredient: the integral over the $\beta$-cycle. However, it is useful to speak more natural language to introduce the $\beta$-cycle into our setup.

\section{Going quantum}

The presented above picture becomes extremely nice and simple if we speak in the language of quantum mechanics. Indeed, the ultimate feature of our system is that we can transform the Riccati equation for $\Phi(t)$ into the Schr\"{o}dinger equation by substitution $\Phi=-\dot{\psi}/(f_{+}\psi)$ and then making gauge transformation $\psi\to\sqrt{f_{+}}\psi$ to remove the first derivative. This Schr\"{o}dinger equation is
\begin{equation}\label{eq:RSJ-Schrod}
    -\ddot{\psi}+V(t)\psi=0,
\end{equation}
where we use original time $t$ and the potential $V(t)$ is given by
\begin{equation}\label{eq:RSJ-Schrod-V}
    V(t)=f_{-}(t)f_{+}(t)+\frac{3}{4}\left(\frac{\dot{f}_{+}(t)}{f_{+}(t)}\right)^2-\frac{\ddot{f}_{+}(t)}{2f_{+}(t)}.
\end{equation}
The eq.~\eqref{eq:RSJ-Schrod} with potential~\eqref{eq:RSJ-Schrod-V} was firstly derived in~\cite{renne1974shunted}. Let us use slow-time $\tau=\omega t$, so our equation becomes
\begin{equation}\label{eq:RSJ-Schrod-U}
    -\omega^2\frac{d^2\psi}{d\tau^2}+\left(V_0(\tau)+\omega^2U(\tau)\right)\psi=0,
\end{equation}
where $V_0(\tau)=f_{-}(\tau)f_{+}(\tau)$. Now we treat $\tau$ as a spatial variable and $\omega$ plays the role of Planck constant. The term $\omega^2U(\tau)$ arises purely from the gauge transformation. We see that the potential $V(\tau)$ consists of two parts, the classical part $V_0(\tau)$ and the quantum-corrected part $\omega^2U(\tau)$. The expression for the classical momentum is
\begin{equation}\label{eq:QM-elliptic-curve}
    p_0(\tau)=\sqrt{-V_0(\tau)}=\sqrt{-f_{-}(\tau)f_{+}(\tau)}.
\end{equation}
The fold point $\tau_0$ becomes exactly the turning point of $V_0(\tau)$. Again, the complexified eq.~\eqref{eq:QM-elliptic-curve} defines the elliptic curve, which is topologically equivalent to the torus with two homology cycles, $\alpha$- and $\beta$-cycles (exactly the same as in our critical curve analysis).

\begin{figure}
    \centering
    \includegraphics[width=\linewidth]{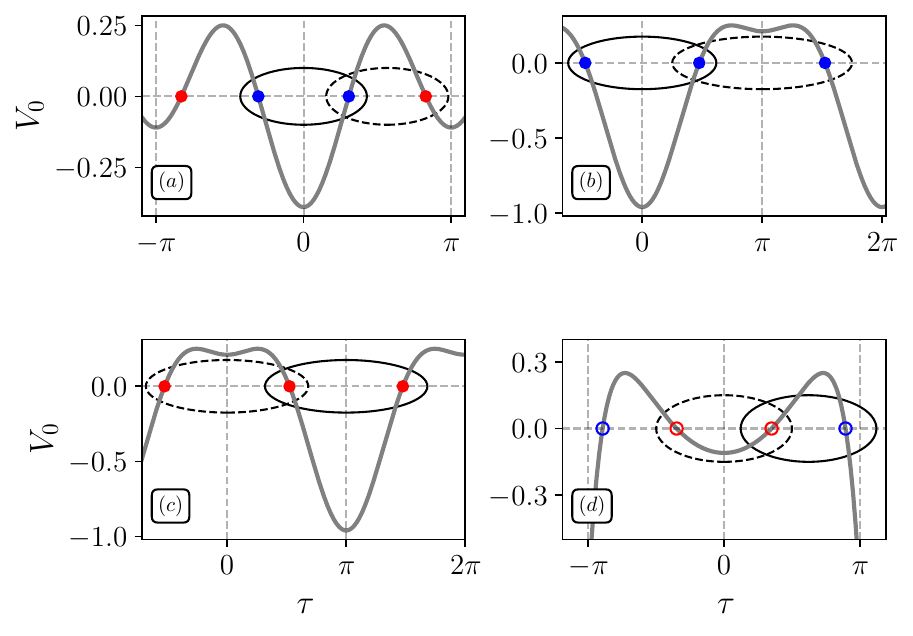}
    \caption{Potential $V_0(\tau)$ in different regimes: (a) 2FP regime: both $\tau_{\pm}\in\mathbb{R}$, (b) SFP regime, $\tau_{-}\in\mathbb{R}$, $\tau_{+}\in\mathbb{C}$, (c) SFP regime: $\tau_{+}\in\mathbb{R}$, $\tau_{-}\in\mathbb{C}$, (d) NFP regime: both fold points are complex (unfilled circles), rotation $\tau\to -i\tau$ was done). Fold point $\tau_{+}$ is colored red, $\tau_{-}$ is colored blue. The $\beta$-cycle is shown by dashed line, the $\alpha$-cycle is shown by solid line.}
    \label{fig:V0-plots}
\end{figure}

Let us plot the potential $V_0(\tau)$ because it will be helpful for further narration, see fig.~\ref{fig:V0-plots} and tab.~\ref{tab:V0-structure}. Notice that when the fold point is complex, the drawn cycle is schematic: it is just a ``shadow'' of cycle in complex plane. Since the total energy in~\eqref{eq:RSJ-Schrod-U} is zero (in fact, we can extract the constant part of the potential, which combines $A$ and $B$, it plays a role of energy), we have the following domains: $V_0(\tau)<0$, where $|B+A\cos\tau|>1$, which is \emph{a classically allowed} region; $V_0(\tau)>0$, when $|B+A\cos\tau|<1$, which is \emph{classically forbidden} region. In the classically allowed region, the monodromy block is \emph{elliptic}, the wavefunction $\psi(\tau)$ is bounded, and it is a \emph{band}. In the classically forbidden region, the monodromy block is \emph{hyperbolic}, the wavefunction is exponentially growing, and it is a \emph{gap}. The non-zero value of $B$ just produces an imbalance between the depth of two wells. Potential $V_0(\tau)$ in different regimes is shown at fig.~\ref{fig:V0-plots}.

\begin{table}[h!]
\begin{tabular}{c|c|c}
                          & regime & $V_0(\tau)$ structure                   \\ \hline
$\tau_{\pm}\in\mathbb{R}$ & 2FP    & $0$- and $\pi$-centered wells \\ \hline
$\tau_{-}\in\mathbb{R}$   & SFP    & $0$-centered well          \\ \hline
$\tau_{+}\in\mathbb{R}$   & SFP    & $\pi$-centered well          \\ \hline
$\tau_{\pm}\in\mathbb{C}$ & NFP    & no real turning points                      
\end{tabular}
\caption{$V_0(\tau)$ in different regimes}
\label{tab:V0-structure}
\end{table}

Now we can say \emph{which} well is quantized. Based on the QM-nature, generically only the deeper well has a level, and the shallow one contributes only through the off-resonant admixture, except the double resonance lines $B=l\omega$, $l\in\mathbb{Z}$.

\begin{table}[h!]
\begin{tabular}{c|c|c}
$\mathcal{M}$ & $\psi$ (QM)       & $\Phi$ (classical)         \\ \hline
hyperbolic    & gap & attr. orbit \\ \hline
elliptic      & band  & no fixed point \\ \hline
parabolic     & band-gap edge         & balanced canard \\
\end{tabular}
\caption{QM-classical correspondence}
\label{tab:QM-classical}
\end{table}

The potential $V(\tau)$ is obviously periodic, so instead of discrete energy levels, the band-gap structure appears. Based on our previous analysis, we have seen that in bands we can touch the unstable branch but immediately jump from it. This is a hint: ducks live somewhere close both to the band and the gap. Precisely at the band-gap edge, something very interesting should happen. Note that in QM set-up, we have a swap of roles; see tab.~\ref{tab:QM-classical} (here we have announced that a band-gap edge corresponds to a balanced canard). 

Last but not least: we virtually always speak about the $A>0$ domain (except for the discussion of the regime chart). Let $A>0$ and we have a single fold point, say $\tau_{+}$. When $A<0$, we still have a real fold point $\tau_{-}=\pi-\tau^{+}$. This causes a shift of integration domains, but the band-gap structure remains the same.

\section{Inspection of duck}

Our hunt is almost over. The QM-based investigation of the bandgap structure was tedious, but it is our hunting riffle. In this section, we discuss the mechanism that causes the existence of canard trajectories, identify maximal canard solutions, provide the relation between the canard trajectory and instanton action, and finally demonstrate the canard trajectories.

Since we refer to the results, in this section we use torus dynamics (i.e. not
Riccati or Schr\"{o}dinger form),
\begin{equation*}
    \dot{\phi}=-\cos\phi + B+A\cos\tau,\quad \dot{\tau}=\omega.
\end{equation*}
Recall that the slow manifold looks as
\begin{equation*}
    \phi_0(\tau)=\pm\arccos\left(B+A\cos\tau\right)
\end{equation*}
with the unstable and stable branches
\begin{equation*}
\begin{gathered}
    \phi_0^{+}=+\arccos\left(B+A\cos\tau\right)\in (0,\pi)\quad\text{(unstable)},\\
    \phi_0^{-}=-\arccos\left(B+A\cos\tau\right)\in (-\pi,0)\quad\text{(stable)}.
\end{gathered}
\end{equation*}
Time reversion exchanges the stability of $\phi_{0}^{+}$ and $\phi_0^{-}$: the
unstable branch is the generic attractor of the backward flow, which is what we
exploit below.

\subsection{All ducks live in instanton?}

The duck that never jumps lives in the centers of QM-bands, and this fact builds a skeleton of the complete picture. Quantum mechanics tells us that the size of QM bands is defined by the aforementioned integral over the $\beta$-cycle. It is known that this integral is nothing more than the doubled instanton action, so
\begin{equation}
    S_{\text{inst}}=\frac{1}{2}\oint_{\beta}d\tau\,\sqrt{V_0(\tau)}
\end{equation}
and the bands are exponentially thin: the $\alpha$-cycle
says \emph{where} a band is (its center is $I_{\alpha}=2\pi\omega(n+1/2)$), and the $\beta$-cycle says \emph{how wide} it is. The width of the band is $\sim\exp(-S_{\text{inst}}/\omega)$.

Since $\omega\to 0$, the characteristic scale $\exp(-S_{\text{inst}}/2\omega)$ is the width of vanishingly thin features in the $(B,A)$-plane. This scale plays a twofold role: it controls the bandwidth, but its second role is much more interesting for our goals. Around the band, in a layer of the same size $\sim\exp(-S_{\text{inst}}/2\omega)$, canard trajectories are what a generic initial condition actually produces! We stress that there are two different thin sets here, and they should not be confused:

\emph{In the parameter space}, the layer of width $e^{-S_{\text{inst}}/2\omega}$ around a band, where canards are displayed by general initial conditions;

\emph{In phase space}, at any parameters inside a gap, a tube of width $e^{-S(\zeta)/\omega}$ around the repelling closed orbit, whose orbits shadow the unstable branch up to the action fraction $S(\zeta)$, where $\zeta$ is \emph{canard fraction} (will be clarified below).

The first set is where all \emph{forward-observable} ducks live; the maximal
duck, which rides the unstable branch across the whole lens, lives in the gap
interior instead and is reachable only through the second one. To understand
this, we need to understand the whole mechanism that provides the existence of
canard trajectory.

\subsection{Local mechanism}

By definition, the canard trajectory implies $O(1)$ time spent on the unstable
branch $\phi_0^{+}$. The only mechanism that allows us to do it is compensation
of accumulated deviation from the unstable branch. This deviation is
\begin{equation*}
    \Delta(\tau,\omega)=\Delta_0(\omega)\exp\left\{+\frac{1}{\omega}\int^{\tau} d\tau'\,\mu(\tau')\right\},
\end{equation*}
so that the total exponent available across a lens is exactly $2S_{\text{inst}}/\omega$, and we can compensate for exponential growth only by
fine-tuning of $\Delta_0(\omega)$. We can do it in a few ways:

\emph{Seed calibration.} We anchor on the unstable branch and integrate the full dynamics backward past the entry fold. This backward construction eliminates the anchor error $O(\omega)$ and allows us to find an initial condition which is a true recessive state with exponential accuracy.

\emph{Parameter tuning.} Move in $(B,A)$-plane to the band-gap edge, staying in the hyperbolic monodromy side in an exponentially thin layer. Here attractive and repulsive orbits are exponentially close.

Note that both ways are related to high-precision computations: for the first case, we have an exponentially small region in the \emph{phase space}, which represents initial conditions that favor canard trajectories; for the second case, we have an exponentially thin region in the \emph{parameter space}, which represents a set of parameters that favor canard trajectories for \emph{general} initial conditions. The cost is the same in both: an $O(1)$ flight of banked action $S(\zeta)$ requires about $0.434\,S(\zeta)/\omega$ decimal digits to be supplied somewhere.

Finally, notice that in the NFP regime (no fold points, $q(\tau)=\pm 1$ has no real solution), the band-gap picture obviously persists, and we still have the $\beta$-cycle integral. But in this regime, there are no canards: the fold points are complex, so there is no place where the flow can cross from the stable to the unstable branch.

\subsection{Global matching}

Now, let us discuss the global picture. We realize both ways to observe canard
trajectories.

\begin{figure}
    \centering
    \includegraphics[width=\linewidth]{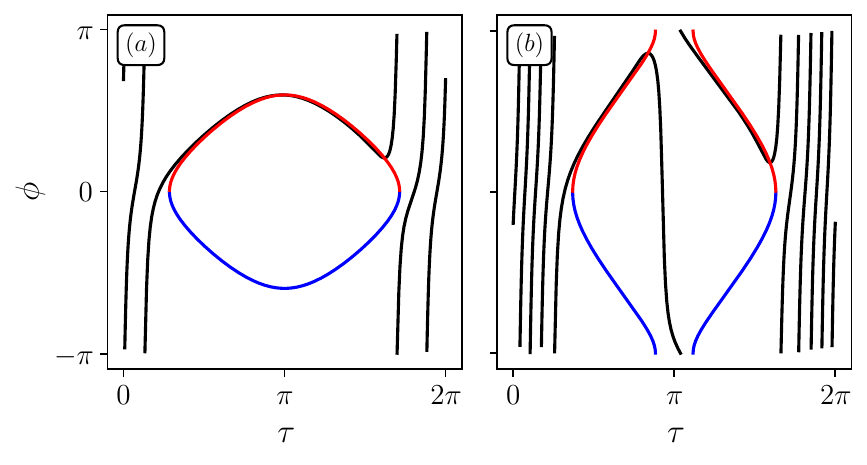}
    \caption{Canard trajectories with $\omega=0.05$ and different parameters:
    (a) $A=1.5$, $B=0.4$ (2FP, two lenses per period,
    $2S_{\text{inst}}/\omega=24.2$ per lens), (b) $A=0.8$, $B=0.5$ (SFP, one
    lens, $2S_{\text{inst}}/\omega=77.1$). Stable branch of slow curve is blue,
    unstable branch is red.}
    \label{fig:canards}
\end{figure}

\emph{Seed calibration.} During the implementation of this strategy, we take four steps. First, we have to select recessive orbit, i.e. kill the forward growing mode. We need to fix four time moments: $\tau_0$ in the running region left of our slow curve, the entry moment $\tau_{L}$, the anchor $\tau_{*}$ on the unstable branch, and $\tau_R$, where our trajectory passes through the right fold point. We choose the parameter $\zeta$, which plays the role of the ``canard fraction''. Our anchor is $\tau_{*}=\tau_L+\zeta(\tau_R-\tau_L)$. Second, perform backward integration with initial condition on the unstable branch, $\phi(\tau_{*})=\arccos(B+A\cos\tau_{*})$ from $\tau_{*}$ to $\tau_L$. This initial condition contains an error of $O(\omega)$, but this error is contracted by the time the solution reaches $\tau_L$. We continue this backward solution through the left fold point to the running region (it is our ``approaching'' part). Third, we calibrate our initial condition: at $\tau_{L}$, we set $\phi(\tau_L)=\phi_{\text{in}}-\Delta_0$, where $\phi_{\text{in}}$ is the value delivered by the backward leg and
\begin{equation}
    \Delta_0(\omega)=\exp\left\{-\frac{1}{\omega}\int_{\tau_L}^{\tau_*}d\tau\mu(\tau)\right\},
\end{equation}
which is the mentioned above contraction. Finally, we start forward integration with our calibrated initial condition, simulate all parts of the trajectory, and glue them. The results are shown in fig.~\ref{fig:canards}. Our compensation is limited: analytically $\Delta_0(\omega)\gtrsim e^{-S_{\text{inst}}/\omega}$, and
numerically $\Delta_0$ cannot go below machine epsilon, so in double precision
\begin{equation}
    \frac{S(\zeta)}{\omega}\lesssim\ln\frac{1}{\varepsilon_{\text{mach}}}\approx 36.
    \label{eq:precision}
\end{equation}

\emph{Parameter tuning.} In fact, we have \emph{three} parameters: $A$, $B$, and $\omega$. The canard condition on the torus is a single scalar condition (it is met on the phase-locking boundaries $|\tr\mathcal{M}|=2$), so no additional parameter beyond these three is needed, and $A$ and $B$ can be fixed. This just reflects the simple idea: for general $A$ and $B$, we can fit $\omega$ so that we will be in a thin layer on the gap side of the band edge. In this regime, we can easily use our computed $\alpha$- and $\beta$-cycle integrals. Again, our compensation ability is bounded by $2S_{\text{inst}}$, therefore, as the balance law below shows, along this route it is in fact bounded by one half of it.

\subsection{Duck that never jumps}

If $B$ and $A$ are such that the well of $V_0(\tau)$ around $\tau=0$ is deep and nearly parabolic, we can forget about the periodicity of $V_0(\tau)$ and work in a single-well approximation. This means that we can expand near the minimum of the potential $V_0(\tau)$ at $\tau=0$,
\begin{equation}
    V_0(\tau)=\frac{1-(A+B)^2}{4}+\frac{A(A+B)}{4}\tau^2+O(\tau^4).
\end{equation}
As should be, we have a simple quantum harmonic oscillator. Since the
Schr\"{o}dinger problem is posed at zero energy, the level condition is
$V_0(0)+\omega\sqrt{c}\,(2n+1)=0$ with $c=A(A+B)/4$, i.e.
\begin{equation}
    \left(A+B\right)^2-1=2\omega\sqrt{A(A+B)}(2n+1),\,n\in\mathbb{Z}_{+}.
    \label{eq:callot}
\end{equation}
Notice that this condition holds only if the well is deep enough to hold the level and if the level sits low in it, i.e. for small $n$ (the rigorous derivation of this simple fact is presented in~\cite{glutsyuk2026parquet}). The anharmonicity of $V_0$ is what limits the accuracy, not the size of $A$ and $B$ themselves. Let us denote $l=(A(A+B)/(4\omega^2))^{1/4}$, then the localized eigenfunctions become
\begin{equation}
    \psi_n(\tau) = C_nH_n\left(l\tau\right)e^{-(l\tau)^2/2},
\end{equation}
where $C_n$ is a normalization constant. Within the single-well approximation
these eigenfunctions are recessive on both sides of the well, and the tunneling between neighboring wells that we have dropped is exactly what will turn the
level into a band.

It is worth noting where \eqref{eq:callot} sits with respect to the band-gap
structure obtained above. It is straightforward to check that our $\alpha$-cycle in the harmonic limit reduces to
\begin{equation}
    \oint_{\alpha}d\tau\,\sqrt{-V_0(\tau)}=\frac{2\pi|V_0(0)|}{\sqrt{A(A+B)}}
\end{equation}
so, the quantization condition reproduces \eqref{eq:callot} identically. In other words, the quantization condition of~\cite{callot1981bifurcations} \emph{is} our $\alpha$-cycle condition evaluated in the harmonic approximation, and this is why its solutions are band centers.

One can ask whether this is a duck at all, since by definition a canard trajectory not only touches the unstable branch but travels across it. It does: the condition ``recessive on both sides of the well'' means, in the language of the flow, that the trajectory rides the unstable branch through the lens, and it does so for an $O(1)$ slow time. What the Hermite function does not show is this flight, because $\psi_n$ is supported in the well, where there is no slow manifold at all, while the flight takes place in the barrier, where $\psi_n$ is exponentially small. This solution is called \emph{duck that never jumps} in~\cite{callot1981bifurcations}; the name refers to the Riccati variable, which does not run through infinity in the lens.

\subsection{Balanced canards}

The band-gap edge, i.e. the region with parabolic monodromy, corresponds to the so-called \emph{balanced} canards. To see why, linearize the torus equation about any periodic orbit and integrate over one period, then the Floquet multiplier $\Lambda$ is given by
\begin{equation}
    \ln\Lambda=\frac{1}{\omega}\oint d\tau\,\mu(\tau)
    =\frac{S_{+}-S_{-}}{\omega},
    \label{eq:balance}
\end{equation}
where $S_{+}$ and $S_{-}$ are the parts of the budget banked on the unstable and stable branch, $S_{+}+S_{-}\simeq
2S_{\text{inst}}$ (the running wells contribute only $O(\omega)$ to
the integral). Hence,
\begin{equation}\label{eq:unstable-bank}
    S_{+}=S_{\text{inst}}+\frac{\omega\ln\Lambda}{2},
\end{equation}
and at the edge $\Lambda=1$, so that our total budget on the unstable branch is
exactly the instanton action,
$S_{+}=S_{-}=S_{\text{inst}}$ and $\Delta_0(\omega)\sim e^{-S_{\text{inst}}/\omega}$. Two consequences are worth stating. Any \emph{stable} periodic orbit has $\ln\Lambda\le 0$, so no forward-observable canard can ever bank more than half of the instanton action on the unstable branch; and deep in a gap $\ln\Lambda$ saturates the envelope, so the repelling closed orbit rides the unstable branch across the whole lens, it is the maximal canard, and it is necessarily unstable.

\begin{figure}
    \centering
    \includegraphics[width=\linewidth]{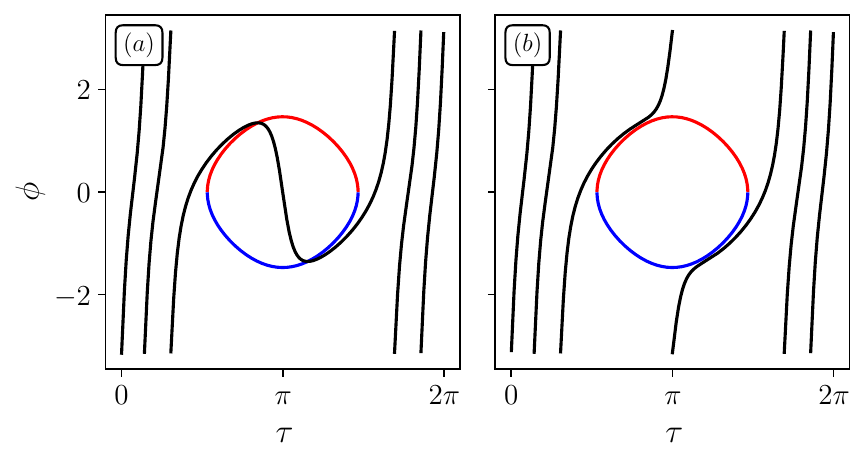}
    \caption{Balanced trajectories with $A=1.0$, $B=1.1$: (a)
    $\omega=0.128476224$, monodromy is parabolic with $\tr\mathcal{M}=+2$; (b)
    $\omega=0.128477263$, monodromy is parabolic with $\tr\mathcal{M}=-2$. In
    both panels the orbit banks exactly one half of the instanton action on the
    unstable branch. Stable branch of slow curve is blue, unstable branch is
    red.}
    \label{fig:balanced-canards}
\end{figure}

Both parabolic points of a band carry a balanced canard, and the two differ by one unit of the winding number $n$: the extra turn is spent as a \emph{head}, i.e. an extra passage of $\phi$ through $\pi$ inside the lens. Since $\sgn\tr\mathcal{M}=(-1)^n$, the two are automatically of opposite trace sign, which is what distinguishes the panels of fig.~\ref{fig:balanced-canards}. The duck that never jumps of the previous section sits between them, at the band center, with the same half budget but without closing into a periodic orbit.

\begin{figure}
    \centering
    \includegraphics[width=\linewidth]{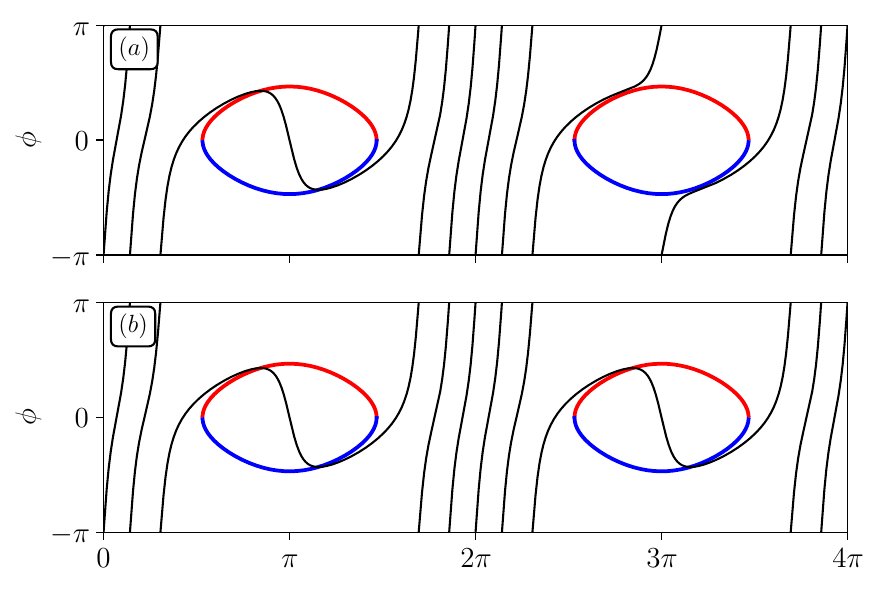}
    \caption{(a) canard trajectory that corresponds to ``duck that never jumps'', (b) balanced canard solution}
    \label{fig:two-types-canards}
\end{figure}

\subsection{Canard explosion}

The two parabolic points bound a band, and by tab.~\ref{tab:QM-classical} the band is precisely the interval where the recurrence map has no fixed point, i.e. the non-locked sliver between the Arnold tongues $\rho=n$ and $\rho=n+1$. Crossing that sliver is the canard explosion. On the tongue side of the left edge, the attractor $\phi_A$ acquires a canard segment and banks $S_+\to S_{\text{inst}}$ by~\eqref{eq:unstable-bank}. At the edge, it annihilates with the repelling canard in a fold of periodic orbits. Inside the band there is no periodic orbit, the rotation number sweeps continuously from $n$ to $n+1$, and where $\tr\mathcal M=0$ the flow carries the duck that never jumps. At
the right edge a new pair is born with one extra unit of winding. The explosion
therefore occupies an interval of width $\sim e^{-S_{\text{inst}}/\omega}$ (SFP;
$\sim e^{-2S_{\text{inst}}/\omega}$ for detuned wells in 2FP). For the RSJ junction it is the riser between consecutive Shapiro steps: the DC voltage moves by $\hbar\omega/2e$ over an exponentially narrow interval of bias current, and every trajectory observed on that riser is a canard.

\section{End of hunt}

\emph{La grande finale}: a duck catalog featuring their habitat ranges, key distinguishing features, and common names. Take a look in tab.~\ref{tab:ducks-hybrid}. One thing we need to do: we have discussed the $\alpha$- and $\beta$-cycles, their roles, and interpretations. We can do more: write down the full and exact quantization condition that will give us the bandgap structure and allow us to take into account higher order $\omega$-corrections.

\begin{table}[t]
\begin{tabular}{l|c|c|c|c|c|c}
canard & locus & $\tr\mathcal{M}$ & $\Lambda$ & $S_{+}/S$ & $n$ & closed \\
\hline
maximal     & gap       & $\gg2$     & $\gg1$      & $1$           & $n$          & yes \\
usual       & near edge & $\gtrsim2$ & $\lesssim1$ & $\lesssim1/2$ & $n$          & yes \\
headless    & edge      & $\pm2$     & $1$         & $1/2$         & $n$          & yes \\
never jumps & centre    & $0$        & ---         & $1/2$         & $n+1/2$ & no  \\
with head\footnotemark[1] & edge & $\mp2$ & $1$ & $1/2$ & $n+1$ & yes \\
\end{tabular}
\footnotetext[1]{Which of the two edges carries the head is parameter
dependent: in the 2FP regime both edges can be headless, the extra unit of
winding being spent in the running well instead.}
\caption{%
Catalog of the canard solutions. $S\equiv 2S_{\text{inst}}$, $\Lambda$ is the
Floquet multiplier and $n$ the winding number over one period; the balance law
$S_{+}/S=(1+\omega\ln\Lambda/S)/2$ relates the last two columns. The
\emph{maximal} duck is the repelling orbit $\phi_{R}$; the \emph{usual} one is
the attractor $\phi_{A}$, which acquires a canard segment only in the layer of
width $\sim e^{-S/2\omega}$ next to an edge. The duck that never jumps is the
overstable solution of Ref.~\cite{callot1981bifurcations}; $\tr\mathcal{M}=0$ gives
$\mathcal{M}^{2}=-\openone$, so it closes only after two periods.}
\label{tab:ducks-hybrid}
\end{table}

\subsection{QM binocular for duck hunting}

The quantum mechanics provides all needed hunting equipment: it automatically resolves the small $\omega$ case, allows to locate ducks, and provides the iterative procedure that is valid on all $\omega$-orders. Let us provide the exact quantization condition. Consider again two integrals over homology cycles of our elliptic curve,
\begin{equation}
\begin{gathered}
    \mathcal{I}_{\alpha}(\omega,A,B)=\oint_{\alpha}d\tau\,p(\tau,\omega),\\ \mathcal{I}_{\beta}(\omega,A,B)=i\oint_{\beta}d\tau\,p(\tau,\omega),
\end{gathered}
\end{equation}
with $p(\tau,\omega)=\sum_{k\geq 0}p_{k}(\tau)\omega^{k}$ is the quantum momentum and $p_0(\tau)=\sqrt{-V_0(\tau)}$. The exact WKB tells us that the centers of QM-bands are given by
\begin{equation}
    \mathcal{I}_{\alpha}(A,B)=2\pi\omega\left(n+\frac{1}{2}\right),\quad n\in\mathbb{Z}_{+},
\end{equation}
and also can tell us that the centers of \emph{QM-gaps} are given by condition $\mathcal{I}_{\alpha}(A,B)=2\pi\omega n$.
The integral $\mathcal{I}_{\beta}(A,B)$ is the contribution of the instanton, feels non-perturbative effects, and tells how wide the $n$-th band is.

The full (and exact) quantization condition is given by
\begin{equation}
\begin{gathered}
    \left|2e^{\gamma}\cos\theta_1\cos\theta_2+\cos\left(\theta_1-\theta_2\right)\right|=1,\\
    \theta_1 = \frac{\re\mathcal{I}_{\alpha}}{2\omega},\quad \theta_2=\theta_1-\frac{\im\mathcal{I}_{\beta}+\pi B}{\omega},\\ \gamma=\frac{2\re\mathcal{I}_{\beta}-\im\mathcal{I}_{\alpha}}{2\omega}.
\end{gathered}
\end{equation}
It is nothing more than the Dillinger-Delabaere-Pham formula (see~\cite{Delabaere1993,Delabaere1997} for original articles and~\cite{sueishi2020quantization} for pedagogical derivation) and we provide its derivation in the Appendix~\ref{app:ddp}. Using the leading $\omega$-order integrals for the $\alpha$- and $\beta$-cycles,
\begin{equation*}
    I_{\alpha}=\oint_{\alpha}d\tau\sqrt{-V_0(\tau)},\quad I_{\beta}=i\oint_{\beta}d\tau\sqrt{-V_0(\tau)},
\end{equation*}
we compute \emph{analytically} (details are given in Appendix) the bandgap structure, see fig.~\ref{fig:bandgap-compar} and also~\ref{fig:bandgap-leading-order}b. Here, we also have taken into account that in the 2FP regime we have double resonances. Notice that the previously obtained Maslov index corresponds just to the case of thick barrier and isolated bands, but it is a good assumption for the location of band centers.

\begin{figure}
    \centering
    \includegraphics[width=\linewidth]{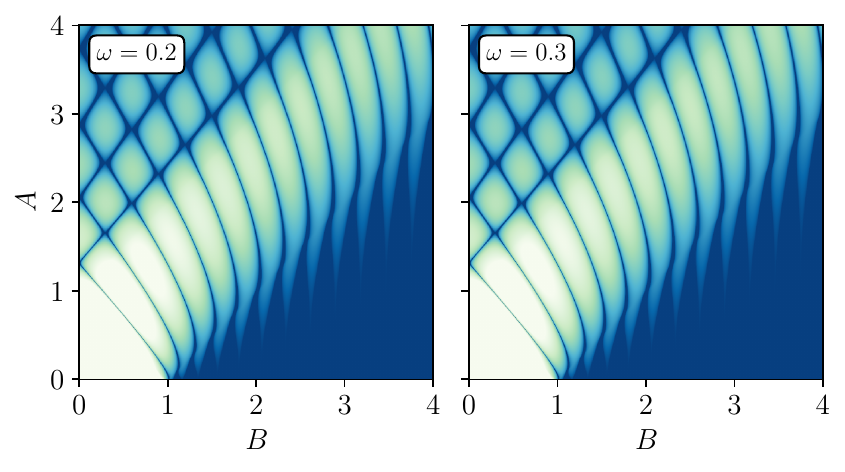}
    \caption{Leading-order band-gap structures for different $\omega$.}
    \label{fig:bandgap-compar}
\end{figure}

Crossing $|A|=1+|B|$, the elliptic modulus of the $\alpha$-cycle vanishes, the holomorphic $\alpha$-period degenerates and $\tau_{+}$ leaves the real axis: the shallow $\pi$-centered well loses its last level, the diatomic lattice becomes monatomic, constrictions cease to exist, and the band width accordingly changes from $\sim e^{-S_{\text{inst}}/\omega}$ (two detuned sublattices, second order in the hopping) to $\sim e^{-S_{\text{inst}}/2\omega}$. On the opposite boundary $|A|=|B|-1$, $|B|>1$ it is the $\beta$-cycle that degenerates: the two folds coalesce, $S_{\text{inst}}\to0$ and $e^{-S_{\text{inst}}/2\omega}\to1$, so bands and gaps become comparable and no fine tuning is needed to see a duck. This is the analog of the instanton condensation~\cite{dunne2017mathieu,basar2017quantumgeometry}, where $\exp[-2\pi\,\mathrm{Im}\,a_0^D/\hbar]\to O(1)$ at the barrier top; on the classical side it is the disappearance of the Arnold tongues, i.e. the loss of Shapiro steps at large DC bias and weak drive.

\subsection{Making all classical again}

The presented above quantization condition is known in the physicists' community, while for people who deal with dynamical systems it is just a mix-up of integrals and geometric stuff. But the tenet ``same equations, same solutions'' works in real life. We just need to compare $\mu_0(\tau)=\pm 2\sqrt{f_{+}(\tau)f_{-}(\tau)}$ and $p_0(\tau)=\sqrt{-f_{-}(\tau)f_{+}(\tau)}$. It is obvious that
\begin{equation*}
    \mu_0(\tau)\,d\tau = \pm 2ip_0(\tau)\,d\tau.
\end{equation*}
This means that the instantaneous Floquet exponent $\mu(\tau,\omega)$ coincides up to the numerical factor $2i$ with the quantum momentum $p(\tau,\omega)$. Since both elliptic curves are isomorphic, i.e. homology cycles coincide, the exact Floquet exponent is related to our integrals over $\alpha$- and $\beta$-cycles, which have already been computed.

Exploiting this correspondence, we see that the following quantization conditions for the exact (in sense of all $\omega$-orders) $\alpha$-cycle integral
\begin{equation}
    \mathcal{I}_{\alpha}=2\pi\omega\left(n+\frac{1}{2}\right),\quad \mathcal{I}_{\alpha}=2\pi\omega n
\end{equation}
defines centers of \emph{classical gaps} (i.e. gaps between Arnold tongues) and centers of the Arnold tongues, respectively. The exact $\beta$-cycle integral $\mathcal{I}_{\beta}$ is related to the width of each \emph{classical gap}, i.e., it completes the picture and allows to obtain the structure of Arnold tongues. At the same time, if the real fold points exist, $\mathcal{I}_{\beta}$ controls the size of the canard window as $e^{-\mathcal{I}_{\beta}/(2\omega)}$ and tells us how to capture the duck.

The leading $\omega$-order picture can be extended by recursive method. The formal expansion $\Phi(\tau,\omega)=\sum_n\omega^n\Phi_n(\tau)$ gives us the recursive relation,
\begin{equation}
    \Phi_n = \frac{1}{2f_{+}(\tau)\Phi_0}\left[\Phi_{n-1}'-f_{+}(\tau)\sum_{k=1}^{n-1}\Phi_k\Phi_{n-k}\right]
\end{equation}
and it is a Gevrey-1 series. In fact, our study is simply the Borel summation of this series. The instantaneous Floquet exponent $\mu(\tau,\omega)=2f_{+}(\tau)\Phi(\tau,\omega)$ can also be expanded into the series. $\mu(\tau,\omega)=\sum_n\mu_n(\tau)\omega^n$. The recursive relation is
\begin{equation}
    \mu_n = \frac{f_{+}(\tau)}{\mu_0}\left(\frac{\mu_{n-1}}{f_{+}(\tau)}\right)'-\frac{1}{2\mu_0}\sum_{k=1}^{n-1}\mu_k\mu_{n-k},
\end{equation}
where $\mu_0=2\sqrt{f_{+}(\tau)f_{-}(\tau)}$ and $n\geq 1$. The $\omega^2$-correction $\mu_2$ has a nice structure:
\begin{equation}
    \mu_2 = \frac{1}{\mu_0}\left\{\int d\tau\mu_0(\tau),\tau\right\}-\frac{1}{\mu_0}\left\{\int d\tau f_{+}(\tau),\tau\right\},
\end{equation}
where $\{\cdot,\cdot\}$ denotes the Schwarzian derivative. The higher order terms contain higher Schwarzians, also known as \emph{Gelfand-Dikii invariants}~\cite{GelfandDikii1975}. As usual for the exact WKB method, the great simplification occurs: only even $\omega$-power terms of the $\mu(\tau,\omega)$ expansion contribute to the integrals.

\section{Discussion and conclusion}

In this paper, we have proposed the direct connection between two different sides of singular perturbation theory. The first one corresponds to the well-known WKB method that appears in quantum mechanics. The second one corresponds to the slow-fast theory of dynamical systems. Despite the fact that this connection is studied intensively (at least recently it was discussed in~\cite{kristiansen2024wkbturning, kristiansen2025wkbeigenvalue, kristiansen2026codim2, kristiansen2026arbitrary}), we have established the direct matching between the existence of canard trajectories and the instanton action in the corresponding quantum-mechanical system. Of course, this connection is strongly relied upon by the M\"{o}bius-type dynamics that gives us the Riccati equation and the resulting \emph{linear} Schr\"{o}dinger equation.

Nevertheless, the theory of dynamical systems knows how to deal with nonlinear problems, too. From our point of view, the best starting point for generalization of our findings is the Melnikov method~\cite{melnikov1963stability}, which allows to take into account non-linear perturbations in Hamiltonian systems. It is known that for certain types of perturbations, exponentially small effects (exponentially small separatrices splitting) take place~\cite{gelfreich1997melnikov,gelfreich2001splitting,guardia2010pendulum}. The handling of these exponentially small effects requires a complexification of time, which is the standard way to deal with instanton trajectories. An alternative route is provided by Treschev's continuous averaging~\cite{treschev1996averaging, treschev1997continuous,treschev2002icm}, a continuous-time version of the Neishtadt averaging procedure~\cite{neishtadt1987delay1} which extracts the exponentially small remainder directly; notably, in the rapidly forced pendulum the naive Melnikov integral is only the leading term of a convergent series, and is not by itself the correct asymptotics.

Of course, one can argue that our observations are highly specific and significant for a fairly small community. But the theory of dynamical systems, especially its ``Hamiltonian'' part, deals with stability loss delay~\cite{neishtadt1987delay1, neishtadt1988delay2} for years without many connections to other fields. Therefore, the proposed connection opens a new point of view for dynamical systems theory and makes us think about a deeper interrelation between quantum mechanics and unusual phenomena in classical physics. Without exaggeration, we can say that at least for certain but quite large class of dynamical systems delay in loss stability is a classical reflection of tunneling phenomenon.

Virtually always realistic systems are affected by noise. It is known that the presence of additive white noise suppresses the existence of canards. The rigorous estimates for the 3D system were obtained in~\cite{berglund2012noisy} and 2D case was considered in~\cite{sowers2008noisycanards}. Including the noise into our set-up is natural, but requires introducing of large deviation principles and analysis of mixing properties in dynamical systems, so it will be presented in a separate study.

It is worth mentioning that since canards play a notable role in neuronal dynamics, the neuromorphic devices like memristors (it was already noticed in~\cite{ginoux2016memristor}), spin-torque nano-oscillators, different micro-electromechanical systems, and even vanilla Josephson junction look prominent for \emph{chasse au canard} and the exponential sensitivity can play a useful role in applications.

\section*{Acknowledgments}

I am very grateful to Alexey Glutsyuk for numerous discussions and sharing the PhD thesis of Jean-Louis Callot. I am also grateful to Dmitrii Treschev for his patience with my non-rigorous statements, mentions of Jean \'{E}calle papers, and clarifications related to separatrices splitting. Finally, I am thankful to Alexander Gorsky for preliminary discussions on the exact WKB method, elliptic curves and instantons. This work was performed within the framework of the state assignment at Steklov Mathematical Institute of Russian Academy of Sciences.

For the numerical solution of differential equations, we use {\julia} programming language and the package \verb|DifferentialEquations|. The handling of elliptic integrals was performed in Python with \verb|scipy| and \verb|mpmath|.

\appendix

\section{Warm-up with systems on torus}\label{app:torus}

Any dynamical system on 2D torus, represented by
\begin{equation}
\begin{cases}
    \dot{\phi}=f_1(\phi,\tau),\\ 
    \dot{\tau}=f_2(\phi,\tau),
\end{cases}
\end{equation}
where $f_{1,2}(\phi,\tau)$ are $2\pi$-periodic in each variable, can be characterized by the Poincar\'{e} rotation number $\rho$, which is given by
\begin{equation}
    \rho=\lim\limits_{t\to\infty}\frac{\phi(t)}{\tau(t)}.
\end{equation}
By definition, the first recurrence map (so-called Poincar\'{e} map) is
\begin{equation}
    \mathcal{P}(\phi_0):S^1_{\phi}\to S^1_{\phi},
\end{equation}
which maps $\phi(0)=\phi_0$ to $\phi(2\pi)$. The Poincar\'{e} rotation number can be computed as the limit,
\begin{equation}
    \rho=\lim\limits_{n\to\infty}\frac{\mathcal{P}^n(\phi_0)-\phi_0}{2\pi n}
\end{equation}
and it is known that, in fact, $\rho$ does not depend on $\phi_0$. If the map $\mathcal{P}$ has a $q$-periodic orbit, then the rotation number is rational, $\rho=p/q\in\mathbb{Q}$. If the map $\mathcal{P}$ has no periodic orbit, then this map represents the rotation by a certain angle $\alpha$ (up to conjugacy, it is Denjoy's theorem) and the rotation number is $\rho=\alpha/(2\pi)$, i.e. it is irrational.

For the M\"{o}bius-type system, the map $\mathcal{P}$ is simply M\"{o}bius transformation $S^1\to S^1$. It can have two fixed points (stable and unstable) on $S^1$, have one neutral fixed point, or have no fixed points. If we have a stable fixed point, then we deal with the periodic orbit $q=1$, so $\rho\in\mathbb{Z}$ here. By \emph{phase-locking}, we mean that under change of parameters $A$ and $B$ in a certain domain, the behavior of $\mathcal{P}$ remains the same: it still has a stable fixed point, so the rotation number is integer in the whole domain.

\section{Elliptic curve analysis}\label{app:ell-curve}

Consider the expression
\begin{equation}
    \Phi^2=\frac{f_{-}(\tau)}{f_{+}(\tau)},\quad f_{\pm}=\frac{1}{2}\pm\frac{B+A\cos\tau}{2}.
\end{equation}
Let us rewrite it with variable $u=\tan(\tau/2)$, so $\cos\tau=(1-u^2)/(1+u^2)$ and we have
\begin{equation}
    \Phi^2=\frac{(1-A-B)+(1+A-B)u^2}{(1+A+B)+(1+B-A)u^2}.
\end{equation}
By definition, it is an elliptic curve. There are four branch points, $\pm u_0$, $\pm u_{\infty}$, where
\begin{equation*}
    u_0^2=\frac{A+B-1}{A-B+1},\quad u_{\infty}^2=\frac{A+B+1}{A-B-1}.
\end{equation*}
We have four branch points, and we deal with double cover. The Riemann-Hurwitz formula says that our elliptic curve is topologically equivalent to the surface of genus $g=4/2-1=1$. To see it, pick two sheets of complex plane, draw two branches $[-u_{\infty},-u_0]$ and $[u_{0},u_{\infty}]$ on each sheet, and finally glue two sheets by these branches.

Now, let us define homology cycles. The $\alpha$-cycle encircles two zeros, i.e. $\pm u_0$. The $\beta$-cycle encircles the points $u_0$ and $u_{\infty}$. This gives the intersection number $\alpha\cdot\beta=1$, as should be. In terms of the original variable $\tau$, the $\alpha$-cycle encircles $\pm\tau_{-}$, the $\beta$-cycle encircles $\tau_{-}$ and $\tau_{+}$, where
\begin{equation}
    \tau_{\pm}=\arccos\left(\frac{\mp 1-B}{A}\right).
\end{equation}
It is worth mentioning that the rescaling $x=u/u_0$ gives the standard Legendre form for our elliptic curve. With $u_{0}$ and $u_{\infty}$, the elliptic curve is
\begin{equation*}
    \Phi^2=C_0\frac{u^2-u_0^2}{u_{\infty}^2-u^2},\quad C_0=\frac{1+u_{\infty}^2}{1+u_{0}^2}.
\end{equation*}
The holomorphic form is given by
\begin{equation*}
    \Omega_{\text{hol}}=\frac{C_0\,du}{\sqrt{(u^2-u_0^2)(u_{\infty}^2-u^2)}}
\end{equation*}
and it defines the torus periods,
\begin{equation*}
    \Omega_{\alpha}=\oint_{\alpha}\Omega_{\text{hol}}=\frac{4C_0K(k)}{u_{\infty}},\,
    \Omega_{\beta}=\oint_{\beta}\Omega_{\text{hol}}=\frac{2iC_0K(k')}{u_{\infty}},
\end{equation*}
with $k^2=u_0^2/u_{\infty}^2$ and $k'=\sqrt{1-k^2}$. Their ratio is the complex modulus of our torus,
\begin{equation*}
    \mathcal{T}=\frac{\Omega_{\beta}}{\Omega_{\alpha}}=\frac{iK(k')}{2K(k)}.
\end{equation*}
The object of our interest is the following differential form, $\Omega=2\sqrt{f_{-}(\tau)f_{+}(\tau)}d\tau$, which can be written as
\begin{equation}
    \Omega=\frac{2C\sqrt{\left(u^2-u_0^2\right)\left(u^2-u_{\infty}^2\right)}}{(1+u^2)^2}\,du,
\end{equation}
where $C=\sqrt{1-(A-B)^2}$. The main difference from the usual case is that our form has poles at $u=\pm i$. Residues at these poles are
\begin{equation}
    \mathop{\mathrm{Res}}_{u=\pm i}\Omega = \pm B.
\end{equation}

Finally, let us discuss the elliptic curve that appears in the QM-based approach. We have
\begin{equation}
    p_0^2=-\frac{(1-B-A\cos\tau)(1+B+A\cos\tau)}{4},
\end{equation}
where $p_0$ is a classical momentum. It is easy to see that the map $(u,\Phi)\mapsto(u,p_0)$ is a biholomorphism, since $1+B+A\cos\tau$ is a rational function of $u$. This means that two elliptic curves are isomorphic to each other. On the $(u,p_0)$-curve, we have the proposed $\Omega$ 1-form up to a numerical coefficient. It is called the action form, or the Seiberg-Witten form, if we want to be fancy. The presence of poles makes our computations more complicated, since we need to combine all three kinds of elliptic integrals.

\section{Maslov index derivation}\label{app:maslov}

Even though it is well known that each simple fold (turning) point gives the $\pi/4$ contribution, we provide the accurate derivation of this fact. Without loss of generality, we can consider the fold point $\tau_{0}=-\tau_{-}$, $\tau_{-}=\arccos((1-B)/A)$, which is a zero of $f_{-}(\tau)$. In slow time our Riccati equation is
\begin{equation}
    \omega\frac{d\Phi}{d\tau}=f_{+}(\tau)\Phi^{2}-f_{-}(\tau).
\end{equation}
Near $\tau_{0}$, we expand $f_{-}(\tau)\approx f_{-}'(\tau_{0})(\tau-\tau_{0})$ and zoom the fold point by setting
\begin{equation}
\begin{gathered}
    \tau-\tau_{0}=\omega^{2/3}\xi,\quad \Phi=\omega^{1/3}\eta, \\
    a=-f_{-}'(\tau_{0})=\tfrac{1}{2}A\sin\tau_{-}>0,\quad f_{+}(\tau_{0})=1,
\end{gathered}
\end{equation} 
where the last equality holds because $f_{+}(\tau)+f_{-}(\tau)=1$. This zooming gives us
\begin{equation}\label{eq:app-normal-form}
    \frac{d\eta}{d\xi}=\eta^{2}+a\xi.
\end{equation}
Its slow curve $\eta_{\pm}=\pm\sqrt{a|\xi|}$ exists only for $\xi\leq 0$, and $\eta_{-}$ is the branch attracting in physical time. Since $\xi$ grows with $\tau$, in the zoomed time $\xi$ the branch $\eta_{-}$ is attracting and $\eta_{+}$ is repelling. We note at once that the slow region lies at $\xi\leq 0$, so this zooming describes the fold at which the trajectory \emph{leaves} the slow curve, exactly as in the main text. The fold at which it is \emph{recaptured} is the mirror image $\tau\to-\tau$, $\Phi\to-\Phi$; in the zoomed variables this is $\xi\to-\xi$, $\eta\to-\eta$, which carries~\eqref{eq:app-normal-form} into $\eta'=\eta^{2}-a\xi$, with the slow region at $\xi\geq 0$ and the labels $\eta_{\pm}$ interchanged. Since $\Phi\to-\Phi$ flips the sign of the linearization exponent $2f_{+}\Phi$, the map also interchanges attracting and repelling branches. Everything below is a property of~\eqref{eq:app-normal-form} and therefore serves both cases.

\begin{figure}
    \centering
    \includegraphics[width=0.5\linewidth]{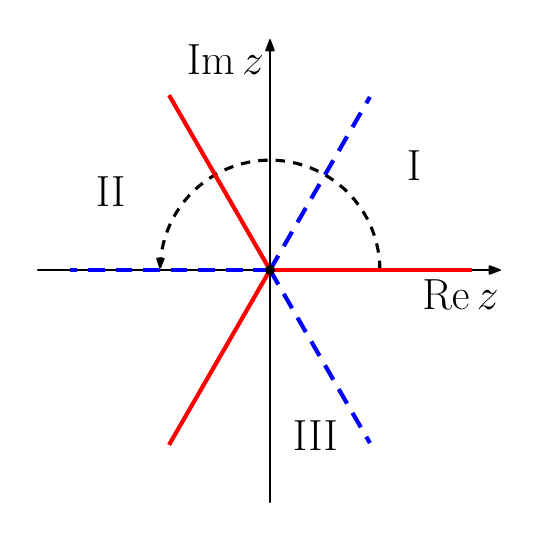}
    \caption{Stokes graph of a simple fold point. Solid rays: the Stokes lines $\arg z=0,\pm 2\pi/3$, on which one of the WKB-solutions is maximally dominant. Dashed rays: the anti-Stokes lines $\arg z=\pi,\pm\pi/3$, on which the two solutions have equal magnitude. We fix the solution by recessiveness along $\arg z=0$ and continue it to $\arg z=\pi$ through the upper half plane, crossing the single Stokes line $\arg z=2\pi/3$.}
    \label{fig:airy-stokes}
\end{figure}

Using substitution $\eta=-\psi'/\psi$, we obtain the Airy equation,
\begin{equation}\label{eq:non-std-Airy}
    \frac{d^{2}\psi}{d\xi^{2}}=-a\xi\psi,
\end{equation}
whose oscillatory region $\xi>0$ is indeed the side on which the slow curve has no real branches. This means that the fold point carries the Airy-type Stokes graph. Note that zeros of $\psi$ are poles of $\eta$; they are harmless, since $\Phi$ lives on $\mathbb{RP}^{1}$. The change of variables $z=-a^{1/3}\xi$ brings eq.~\eqref{eq:non-std-Airy} to the standard form,
\begin{equation}\label{eq:std-Airy}
    \frac{d^{2}\psi}{dz^{2}}=\psi z ,
\end{equation}
and maps the slow region $\xi\leq 0$ onto the ray $\arg z=0$ and the oscillatory region $\xi\to+\infty$ onto the ray $\arg z=\pi$. The WKB-solutions are
\begin{equation}
    \psi^{\pm}=z^{-1/4}\exp\left\{\pm\frac{2z^{3/2}}{3}\right\}.
\end{equation}
By definition, the Stokes lines are given by
\begin{equation}
    \im\int_{0}^{z}dz\,\sqrt{z}=\im\frac{2z^{3/2}}{3}=0,
\end{equation}
so they are three rays $\arg z=0,\pm 2\pi/3$, emanating from $z=0$, see fig.~\ref{fig:airy-stokes}. They cut the plane into three sectors, and inside each sector the coefficients of the WKB-solutions are constant. Since $\re z^{3/2}=|z|^{3/2}\cos(3\arg z/2)$, the solution $\psi^{+}$ is dominant on
the ray $\arg z=0$ and recessive on the rays $\arg z=\pm 2\pi/3$; on the three anti-Stokes lines $\arg z=\pi,\pm\pi/3$ the two solutions have equal magnitude, so the solution is oscillatory there. We write the full solution of~\eqref{eq:std-Airy} as a linear combination $\psi=c^{+}\psi^{+}+c^{-}\psi^{-}$, with the coefficients $(c_{\mathrm{I}}^{\pm})$ in the sector $\mathrm{I}=\{0<\arg z<2\pi/3\}$ and $(c_{\mathrm{II}}^{\pm})$ in the sector $\mathrm{II}=\{2\pi/3<\arg z<4\pi/3\}$, which contains the negative real axis.

Let us pick the solution which is recessive along the ray $\arg z=0$, that is, $c_{\mathrm{I}}^{+}=0$, and normalize $c_{\mathrm{I}}^{-}=1$. Such a condition is well posed precisely because $\psi^{+}$ is dominant on this ray: a condition on the coefficient of the recessive solution would not be, and this is the whole content of the Stokes phenomenon. Indeed, decay along $\arg z=0$ means $(d\psi/dz)/\psi\sim-z^{1/2}$, and since
\begin{equation}\label{eq:eta-chain}
    \eta=-\frac{1}{\psi}\frac{d\psi}{d\xi}
    =a^{1/3}\,\frac{1}{\psi}\frac{d\psi}{dz},
\end{equation}
this gives $\eta\sim-\sqrt{a|\xi|}$, and we touch the branch $\eta_{-}$. In the present
orientation this is the branch attracting in physical time, so the selected solution is simply the trajectory that has just been released by the fold; the actual trajectory differs from it by $O(e^{-2S/\omega})$, $S=\int\sqrt{f_{+}f_{-}}\,d\tau$, which does not affect the phase in the leading order. At the mirror fold $+\tau_{-}$, where the trajectory is captured, the same solution is a canard. The derivation is insensitive to which of the two we have in mind.

Now we continue $\psi$ from $\arg z=0$ to $\arg z=\pi$ through the upper half plane. On this route, we cross the single Stokes line $\arg z=2\pi/3$, on which $\psi^{-}$ is dominant and $\psi^{+}$ is recessive. The jump therefore occurs in the coefficient of the \emph{recessive} solution, because a multiple of it may be added invisibly against the dominant background:
\begin{equation}
    \begin{pmatrix}
        c_{\mathrm{II}}^{+} \\ c_{\mathrm{II}}^{-}
    \end{pmatrix}
    =
    \begin{pmatrix}
        1 & i \\ 0 & 1
    \end{pmatrix}
    \begin{pmatrix}
        c_{\mathrm{I}}^{+} \\ c_{\mathrm{I}}^{-}
    \end{pmatrix}
    =
    \begin{pmatrix}
        i \\ 1
    \end{pmatrix}.
\end{equation}
The Stokes constant $i$ is fixed by the reality of~\eqref{eq:std-Airy}: the continuation through the lower half plane gives $c^{+}=-i$ together with the other branch of $z^{-1/4}$, and both routes must produce the same function. So, in the sector $\mathrm{II}$ we have $\psi=i\psi^{+}+\psi^{-}$. And it is completely correct: on the ray $\arg z=\pi$ the two exponentials have equal magnitude, so we expect an oscillatory solution there. We should also take into account the continuation of $z^{-1/4}$ and of $z^{3/2}=-i|z|^{3/2}$, and write on this ray
\begin{equation}
    \psi^{\pm}=|z|^{-1/4}e^{-i\pi/4}e^{\mp i\theta},\quad
    \theta=\frac{2|z|^{3/2}}{3}.
\end{equation}
The oscillatory solution is
\begin{equation}\label{eq:app-osc}
    \psi=2|z|^{-1/4}\cos\left(\frac{2|z|^{3/2}}{3}-\frac{\pi}{4}\right),
\end{equation}
which is nothing but $2\sqrt{\pi}\,\mathrm{Ai}(z)$.

It remains to return to the trajectory. Since $|z|=a^{1/3}\xi$ for $\xi>0$, the phase $\theta$ coincides with that of the main text,
\begin{equation}
    \theta=\frac{2|z|^{3/2}}{3}=\frac{2\sqrt{a}}{3}\xi^{3/2},
\end{equation}
and $d\theta/dz=-|z|^{1/2}$ for $z<0$. Together with~\eqref{eq:eta-chain}, the solution~\eqref{eq:app-osc} gives for $\xi\to+\infty$
\begin{equation}
    \eta(\xi)\sim\sqrt{a\xi}\,\tan\left(\theta-\frac{\pi}{4}\right),
\end{equation}
which is precisely eq.~\eqref{eq:eta-asym} of the main text. Equivalently, in terms of the standard Airy function, $\eta=a^{1/3}\mathrm{Ai}'(z)/\mathrm{Ai}(z)$ with $z=-a^{1/3}\xi$, i.e.\ eq.~\eqref{eq:eta0}. The $-\pi/4$ is carried by the WKB-solution $\psi$, while $\eta$, being a logarithmic derivative, is a tangent of it. This is the announced $\pi/4$ contribution of a simple fold point.

\section{Elliptic integrals computation}\label{app:ell-integrals}

The integrals over $\alpha$- and $\beta$-cycles with our differential are directly related to the usual elliptic integrals. They can be evaluated explicitly in terms of complete elliptic integrals using the using the tables of Byrd and Friedman~\cite{ByrdFriedman1971} (hereafter BF). Generally, we deal with integrals of form
\begin{equation*}
    \int \frac{du\,\sqrt{P(u)}}{Q(u)},
\end{equation*}
where the Weierstrass change $u=\tan(\tau/2)$ is used. Polynomials $P(u)$ and $Q(u)$ are
\begin{equation*}
    P(u)=\left(u_0^2-u^2\right)\left(u_{\infty}^2-u^2\right),\,Q(u)=\left(1+u^2\right)^2.
\end{equation*}
Moving across different regimes in the $(B,A)$-plane, the roots $u_0$ and $u_{\infty}$ can be real or purely imaginary. The most complicated case corresponds to purely imaginary $u_0$ and $u_{\infty}$ such that the double poles at $u=\pm i$ are enclosed by an integration contour. We have four regimes (2FP, SFP, NFP-diamond, NFP-tail), so naively we need to compute 8 integrals. Fortunately, we can avoid this by careful continuation from 2FP regime to all other regimes across domain wall lines $A=B+1$ and $A=B-1$.

The continuation is based on the regime switching, i.e. transformation of the elliptic modulus. We use the following transformations:
\begin{table}[h!]
\centering
\begin{tabular}{c|c}
$k_{\alpha}^2+k_{\beta}^2=1$  & S-duality \\ \hline
$k^2\to -k^2/(1-k^2)$    & T-duality \\ \hline
$k^2\to 1/k^2$                & Dehn twist \\
\end{tabular}
\end{table}

Switching to imaginary modulus, i.e. T-duality transformation, can be written as
\begin{equation*}
    E(k)=\sqrt{1-k^2}E(k'),\,K(k)=\frac{K(k')}{\sqrt{1-k^2}}
\end{equation*}
\begin{multline*}
    \Pi(n,k)=\frac{k^2}{(k^2-n)\sqrt{1-k^2}}K(k')+\\+\frac{n}{(n-k^2)\sqrt{1-k^2}}\Pi\left(\frac{n-k^2}{1-k^2},k'\right),
\end{multline*}
where $(k')^2=-k^2/(1-k^2)$.
Switching to reciprocal modulus, i.e. the Dehn twist, is given by relations
\begin{align*}
E(k) &= \begin{multlined}[t]
        kE(\kappa)-\frac{k^2-1}{k}K(\kappa)\pm \\
        \pm ik\left(E(\sqrt{1-\kappa^2})-K(\sqrt{1-\kappa^2})/k^2\right),
       \end{multlined} \\
K(k) &= \frac{K(\kappa)}{k}\mp\frac{i}{k}K\left(\sqrt{1-\kappa^2}\right), \\
\Pi(n,k) &= \frac{1}{k}\Pi\left(\frac{n}{k^2},\kappa\right)
          \mp\frac{i}{k(1-n)}\Pi\left(-\frac{n}{(1-n)(k')^2},\frac{1}{k'}\right).
\end{align*}
As we will see further, in the 2FP regime, the elliptic modulus is given by
\begin{equation*}
    k^2=\frac{u_0^2}{u_{\infty}^2}=\frac{(A-1)^2-B^2}{(A+1)^2-B^2}\in (0,1)
\end{equation*}
and precisely on the line $A=B+1$ it vanishes to zero. Below the line $A=B+1$, it becomes negative, so deal with imaginary modulus switching, i.e. T-duality. In the SFP regime strips, the new modulus $k'$ is
\begin{equation}
    \left(k'\right)^2=\frac{B^2-(A-1)^2}{4A}\in(0,1),
\end{equation}
and on the line $A=B-1$ this quantity becomes equal to $1$, so we have reciprocal modulus switching, i.e. the Dehn twist.

Our computation strategy is following: compute integrals in 2FP domain, then use regime switching formulas for other domains, and pay attention to the pole contribution. We also checked our answers by direct evaluation of the integrals in each regime and by numerical integration. In numerics, we have used the \verb|mpmath| and \verb|scipy| libraries. For visualization, vectorization based on \verb|numpy| was performed and the elliptic integral of the third kind was realized with the help of the Carlson symmetric forms, which are available in \verb|scipy|.

\subsection*{2FP regime: \texorpdfstring{$\alpha$-cycle}{alpha-cycle} integral}

When both fold points are real, the $\alpha$-cycle lives purely on the real $u$-axis. In terms of the $\tau$-variable, we have
\begin{multline}
    I_{\alpha}=2\int_{-\tau_{-}}^{\tau_{-}}d\tau\sqrt{-V_0(\tau)}=2\int_{-\tau_{-}}^{\tau_{-}}d\tau\sqrt{-f_{-}(\tau)f_{+}(\tau)}=\\=\int_{-\tau_{-}}^{\tau_{-}}d\tau\sqrt{(B+A\cos\tau)^2-1}.
\end{multline}
Therefore, by symmetry, we can write $I_{\alpha}=2I$, where the integral $I$ is given by
\begin{equation}
    I=\int_{0}^{\tau_{-}}\sqrt{(B+A\cos\tau)^2-1}.
\end{equation}
Using the Weierstrass substitution $u=\tan(\tau/2)$, we obtain the following,
\begin{equation}
    I=C\int_{0}^{u_0}du\frac{\sqrt{(u_0^2-u^2)(u_{\infty}^2-u^2)}}{(1+u^2)^2}.
\end{equation}
where $C=2\sqrt{(A-B)^2-1}>0$ and $0<u_0<u_{\infty}$. The change $u=u_0\sin\phi$ gives us
\begin{equation}
    I=Cu_{\infty}u_0^2\int_{0}^{\pi/2}\frac{d\phi\sqrt{1-k^2\sin^2\phi}\cos^2\phi}{\left(1+u_0^2\sin^2\phi^2\right)^2},
\end{equation}
with $k=u_0/u_{\infty}$. The last integral is decomposable to combination of complete elliptic integrals with coefficients. Using the definitions of complete elliptic integrals with substitutions from BF \S 210 and \S 254, \S 256, \S 258, we obtain
\begin{multline}
    I=\sqrt{(A+1)^2-B^2}E(k)-\\-\frac{2(B+1)(A-B+1)}{\sqrt{(A+1)^2-B^2}}K(k)+\\+
    \frac{4AB}{\sqrt{(A+1)^2-B^2}}\Pi(-n,k),
\end{multline}
where coefficients $k$ and $n$ are
\begin{equation}\label{eq:k-alpha}
    k^2=\frac{(A-1)^2-B^2}{(A+1)^2-B^2},\quad n=\frac{A+B-1}{A-B+1}
\end{equation}
and $k\in(0,1)$. Here $\Pi(n,k)$ is the third kind complete elliptic integral,
\begin{equation*}
    \Pi(n,k)=\int_{0}^{\pi/2}\frac{d\phi}{(1-n\sin^2\phi)\sqrt{1-k^2\sin^2\phi}}.
\end{equation*}

\subsection*{2FP regime: \texorpdfstring{$\beta$-cycle}{beta-cycle} integral}

In the 2FP regime, the $\beta$-cycle integral looks as
\begin{multline}
    I_{\beta}=2\int_{\tau_{-}}^{\tau_{+}}d\tau\sqrt{V_0(\tau)}=2\int_{\tau_{-}}^{\tau_{+}}d\tau\sqrt{f_{-}(\tau)f_{+}(\tau)}=\\=\int_{\tau_{-}}^{\tau_{+}}d\tau\sqrt{1-(B+A\cos\tau)^2}.
\end{multline}
In the $u$-variable, we have
\begin{equation}
    I_{\beta}=C\int_{u_0}^{u_{\infty}}du\frac{\sqrt{(u^2-u_0^2)(u_{\infty}^2-u^2)}}{(1+u^2)^2},
\end{equation}
where again $C=2\sqrt{(A-B)^2-1}>0$ and we have taken into account that $u$ runs from $u_0$ to $u_{\infty}$ with $0<u_0<u_{\infty}$, so $u_{0}^2-u^2<0$. Since $u$ now varies between the two positive branch points, the
substitution used for the $\alpha$-cycle no longer applies; the
appropriate is
\begin{equation}
    u^2=u_{\infty}^2-(u_{\infty}^2-u_0^2)\sin^2\phi,
\end{equation}
which gives
\begin{multline}
    I_{\beta}=\frac{C\,(u_{\infty}^2-u_0^2)^2}{u_{\infty}(1+u_{\infty}^2)^2}\int_{0}^{\pi/2}\frac{d\phi\,\sin^2\phi\cos^2\phi}
    {(1-n\sin^2\phi)^2\sqrt{1-k^2\sin^2\phi}},
\end{multline}
with $k$ and $n$ are
\begin{equation}\label{eq:k-beta}
    k^2=\frac{4A}{(A+1)^2-B^2},\quad n=\frac{2}{A-B+1}.
\end{equation}
Again, the last integral can be decomposed into the combination of complete elliptic integrals. Notice that S-duality holds: $k_{\beta}^2+k_{\alpha}^2=1$, where $k_{\beta}^2$ is eq.~\eqref{eq:k-beta} and $k_{\alpha}$ is eq.~\eqref{eq:k-alpha}.

The explicit expressions for the $\alpha$- and $\beta$-cycle related integrals in SFP and NFP regimes can be obtained by the mentioned transformations (T-duality, Dehn twist). The branch choice is fixed by the constraint that the $\alpha$-cycle integral must be complex in the NFP regime.

\section{DDP quantization condition}\label{app:ddp}

Recall that we deal with the Schr\"{o}dinger equation,
\begin{equation}\label{eq:RSJ-Schrod-app}
    -\omega^2\psi''+V(\tau)\psi=0,\,V(\tau)=V_0(\tau)+\omega^2U(\tau),
\end{equation}
where $V_0(\tau)=f_{-}(\tau)f_{+}(\tau)$ and $U(\tau)$ is given by
\begin{equation*}
    U(\tau)=\frac{3}{4}\left(\frac{f_{+}'(\tau)}{f_{+}(\tau)}\right)^2-\frac{f_{+}''(\tau)}{2f_{+}(\tau)}.
\end{equation*}
Instead of direct work with Schr\"{o}dinger equation, it is convenient to use the linear system (which is regular),
\begin{equation}\label{eq:RSJ-regular}
\omega\frac{dY}{d\tau}=C(\tau)Y,\,
C=\begin{pmatrix}0&f_+\\f_-&0\end{pmatrix},
\,Y=\begin{pmatrix}x \\ y\end{pmatrix}.
\end{equation}
Where $f_{+}(\tau)\neq 0$, we eliminate $y$-variable and set $x=\sqrt{f_{+}}\psi$, which again gives~\eqref{eq:RSJ-Schrod-app}. If $f_{+}(\tau)=0$, we use~\eqref{eq:RSJ-regular}. Let $F(0)=I$ be its fundamental matrix, so the monodromy is $\mathcal{M}=F(2\pi)$. Since $C(\tau)$ is traceless, we have $\det\mathcal{M}=1$. Let $D=\tr\mathcal{M}/2$ (Floquet or Hill discriminant), in gaps we have $|D|>1$, in bands $|D|<1$, and the bandgaps are $D=\pm 1$.

In the 2FP regime, both folds points are real. Using the classical actions $I_{\alpha}$ and $I_{\beta}$, we define the well phases, $\theta_1=I_{\alpha}/(2\omega)$ and $\theta_2=(I_{\alpha}-2\pi B)/(2\omega)$, and the barrier exponent, $\gamma=I_{\beta}/(2\omega)$. With a common Wronskian normalization, the modes rotate in opposite directions in the two wells because $f_+$ changes sign. Writing $R(\theta)=\operatorname{diag}(e^{i\theta},e^{-i\theta})$, the standard Airy connection model gives us
\begin{equation}
    \mathcal{M}_{\text{A}}=H_\gamma R(\theta_0)H_\gamma R(-\theta_\pi),
\end{equation}
where the Airy phase offsets are absorbed into the mode convention, and we have denoted
\begin{equation}
    H_\gamma=\begin{pmatrix}c&-is\\is&c\end{pmatrix}
\end{equation}
where $c^2-s^2=1$, $c=e^{\gamma}+e^{-\gamma}/4$, and $s=e^{\gamma}-e^{-\gamma}/4$.
Computing the trace, we find
\begin{equation}\label{eq:RSJ-2FP-discriminant}
    D_{\rm A}=\cos(\theta_1-\theta_2)+2s^2\cos\theta_1\cos\theta_2,
\end{equation}
and the WKB edges satisfy $D_{\rm A}=\pm1$, with $s^2\sim e^{2\gamma}$, so we obtain the proposed quantization condition. Away from merging folds, $D-D_{\rm A}=O(\omega e^{2\gamma})$, so leading WKB alone does not resolve exponentially narrow edges. The term $\omega^2U$ is essential for this connection. Near the point $\tau_+$, we have $U\sim 3/[4(\tau-\tau_+)^2]$, so $\omega^2U$ and $V_0$ are comparable on the Airy scale $|\tau-\tau_+|=O(\omega^{2/3})$. Dropping the $U$-correction will change the first cosine in \eqref{eq:RSJ-2FP-discriminant} to $\cos(\theta_0+\theta_\pi)$, which produces the wrong leading-order structure. At the line $A=B+1$, the $\pi$-centered well shrinks and the $\beta$-cycle integral obtains the imaginary contribution and in the SFP regime we have $2\pi B=\re\mathcal{I}_{\alpha}-2\im\mathcal{I}_{\beta}$, so we have to add $-\im\mathcal{I}_{\beta}$ contribution to $\theta_2$.

\bibliography{refs}

@article{benoit1981chasse,
  author  = {Beno\^{\i}t, {\'E}ric and Callot, Jean-Louis and Diener, Francine and Diener, Marc},
  title   = {Chasse au canard},
  journal = {Collectanea Mathematica},
  volume  = {31--32},
  number  = {1--3},
  pages   = {37--119},
  year    = {1981}
}

@phdthesis{callot1981bifurcations,
  author  = {Callot, Jean-Louis},
  title   = {Bifurcations du portrait de phase pour des \'{e}quations diff\'{e}rentielles lin\'{e}aires du second ordre ayant pour type l'\'{e}quation d'Hermite},
  school  = {Universit\'{e} Louis Pasteur, Strasbourg},
  year    = {1981},
  address = {Strasbourg},
  note    = {Institut de Recherche Math\'{e}matique Avanc\'{e}e (IRMA); defended 24 June 1981},
  type    = {Th\`{e}se}
}

@article{benoit1983systemes,
  author  = {Beno\^{\i}t, {\'E}ric},
  title   = {Syst{\`e}mes lents--rapides dans $\mathbb{R}^3$ et leurs canards},
  journal = {Ast{\'e}risque},
  volume  = {109--110},
  pages   = {159--191},
  year    = {1983}
}

@incollection{eckhaus1983relaxation,
  author    = {Eckhaus, Wiktor},
  title     = {Relaxation oscillations including a standard chase on {F}rench ducks},
  booktitle = {Asymptotic Analysis II},
  series    = {Lecture Notes in Mathematics},
  volume    = {985},
  pages     = {449--494},
  year      = {1983},
  publisher = {Springer},
  doi       = {10.1007/BFb0062381}
}

@article{callot1985visibles,
  author  = {Callot, Jean-Louis},
  title   = {Solutions visibles de l'{\'e}quation de Schr{\"o}dinger},
  journal = {Math{\'e}matiques finitaires et analyse non standard},
  pages   = {105--119},
  year    = {1985}
}

@article{callot1993champs,
  author  = {Callot, Jean-Louis},
  title   = {Champs lents-rapides complexes \`{a} une dimension lente},
  journal = {Annales scientifiques de l'\'{E}cole Normale Sup\'{e}rieure},
  volume  = {26},
  pages   = {149--173},
  year    = {1993},
  doi     = {10.24033/asens.1669},
  note    = {S\'{e}rie 4}
}

@article{dumortier1996cycles,
  author    = {Dumortier, Freddy and Roussarie, Robert},
  title     = {Canard Cycles and Center Manifolds},
  journal   = {Memoirs of the American Mathematical Society},
  volume    = {121},
  number    = {577},
  year      = {1996},
  publisher = {American Mathematical Society},
  doi       = {10.1090/memo/0577}
}

@book{de2021canard,
  title={Canard cycles: from birth to transition},
  author={De Maesschalck, Peter and Dumortier, Freddy and Roussarie, Robert H},
  volume={73},
  year={2021},
  publisher={Springer},
  doi={10.1007/978-3-030-79233-6}
}

@article{tikhonov1952systems,
  author  = {Tikhonov, Andrey N.},
  title   = {Systems of differential equations containing small parameters in the derivatives},
  journal = {Matematicheskii Sbornik},
  volume  = {31(73)},
  number  = {3},
  pages   = {575--586},
  year    = {1952}
}

@article{fenichel1979gspt,
  author  = {Fenichel, Neil},
  title   = {Geometric singular perturbation theory for ordinary differential equations},
  journal = {Journal of Differential Equations},
  volume  = {31},
  number  = {1},
  pages   = {53--98},
  year    = {1979},
  doi     = {10.1016/0022-0396(79)90152-9}
}

@incollection{jones1995gspt,
  author    = {Jones, Christopher K. R. T.},
  title     = {Geometric singular perturbation theory},
  booktitle = {Dynamical Systems (Montecatini Terme, 1994)},
  series    = {Lecture Notes in Mathematics},
  volume    = {1609},
  pages     = {44--118},
  year      = {1995},
  publisher = {Springer},
  address   = {Berlin, Heidelberg},
  doi       = {10.1007/BFb0095239}
}

@book{kuehn2015multiscale,
  author    = {Kuehn, Christian},
  title     = {Multiple Time Scale Dynamics},
  series    = {Applied Mathematical Sciences},
  volume    = {191},
  year      = {2015},
  publisher = {Springer},
  address   = {Cham},
  doi       = {10.1007/978-3-319-12316-5}
}

@incollection{jardonkojakhmetov2021blowup,
  author        = {Jard{\'o}n-Kojakhmetov, Hildeberto and Kuehn, Christian},
  editor        = {Galaz-Garc{\'i}a, Fernando and Gonz{\'a}lez-Tokman, Cecilia and Pardo Mill{\'a}n, Juan Carlos},
  title         = {A Survey on the Blow-Up Method for Fast-Slow Systems},
  booktitle     = {Mexican Mathematicians in the World},
  series        = {Contemporary Mathematics},
  volume        = {775},
  pages         = {115--160},
  year          = {2021},
  publisher     = {American Mathematical Society},
  address       = {Providence, RI},
  doi           = {10.1090/conm/775/15591},
  eprint        = {1901.01402},
  archivePrefix = {arXiv},
  primaryClass  = {math.DS}
}

@article{krupa2001fold,
  author  = {Krupa, Martin and Szmolyan, Peter},
  title   = {Extending Geometric Singular Perturbation Theory to Nonhyperbolic Points---Fold and Canard Points in Two Dimensions},
  journal = {SIAM Journal on Mathematical Analysis},
  volume  = {33},
  number  = {2},
  pages   = {286--314},
  year    = {2001},
  doi     = {10.1137/S0036141099360919}
}

@article{krupa2001relaxation,
  author  = {Krupa, Martin and Szmolyan, Peter},
  title   = {Relaxation Oscillation and Canard Explosion},
  journal = {Journal of Differential Equations},
  volume  = {174},
  number  = {2},
  pages   = {312--368},
  year    = {2001},
  doi     = {10.1006/jdeq.2000.3929}
}

@article{desroches2011curvature,
  author    = {Desroches, M. and Jeffrey, M. R.},
  title     = {Canards and curvature: the ``smallness of {$\epsilon$}`` in slow-fast dynamics},
  journal   = {Proceedings of the Royal Society A: Mathematical, Physical and Engineering Sciences},
  volume    = {467},
  number    = {2132},
  pages     = {2404--2421},
  year      = {2011},
  publisher = {The Royal Society Publishing},
  doi       = {10.1098/rspa.2011.0053}
}

@article{desroches2012mmo,
  author  = {Desroches, Mathieu and Guckenheimer, John and Kuehn, Christian and Krauskopf, Bernd and Osinga, Hinke M. and Wechselberger, Martin},
  title   = {Mixed-Mode Oscillations with Multiple Time Scales},
  journal = {SIAM Review},
  volume  = {54},
  number  = {2},
  pages   = {211--288},
  year    = {2012},
  doi     = {10.1137/100791233}
}

@article{melnikov1963stability,
  author  = {Mel'nikov, V. K.},
  title   = {On the stability of the center for time periodic perturbations},
  journal = {Transactions of the Moscow Mathematical Society},
  volume  = {12},
  pages   = {1--57},
  year    = {1963}
}

@article{gelfreich1997melnikov,
  author    = {Gelfreich, V.G},
  title     = {Melnikov method and exponentially small splitting of separatrices},
  journal   = {Physica D: Nonlinear Phenomena},
  volume    = {101},
  number    = {3-4},
  pages     = {227-248},
  year      = {1997},
  publisher = {Elsevier BV},
  doi       = {10.1016/s0167-2789(96)00133-9}
}

@article{gelfreich2001splitting,
  author    = {Gelfreich, V G and Lazutkin, V F},
  title     = {Splitting of separatrices: perturbation theory and exponential smallness},
  journal   = {Russian Mathematical Surveys},
  volume    = {56},
  number    = {3},
  pages     = {499-558},
  year      = {2001},
  publisher = {Steklov Mathematical Institute},
  doi       = {10.1070/rm2001v056n03abeh000394}
}

@article{guardia2010pendulum,
  author    = {Guardia, Marcel and Oliv{\'e}, Carme and Seara, Tere M.},
  title     = {Exponentially Small Splitting for the Pendulum: A Classical Problem Revisited},
  journal   = {Journal of Nonlinear Science},
  volume    = {20},
  number    = {5},
  pages     = {595-685},
  year      = {2010},
  publisher = {Springer Science and Business Media LLC},
  doi       = {10.1007/s00332-010-9068-8}
}

@misc{kristiansen2026codim2,
  author        = {Kristiansen, K. U.},
  title         = {A geometric approach to exponentially small splitting: The generic zero-Hopf bifurcation of co-dimension two},
  year          = {2026},
  doi           = {10.48550/arXiv.2603.12103},
  eprint        = {2603.12103},
  archivePrefix = {arXiv},
  primaryClass  = {math.DS}
}

@misc{kristiansen2026arbitrary,
  author        = {Kristiansen, K. U.},
  title         = {A geometric approach to exponentially small splitting: Zero-Hopf bifurcations of arbitrary co-dimension},
  year          = {2026},
  doi           = {10.48550/arXiv.2603.12115},
  eprint        = {2603.12115},
  archivePrefix = {arXiv},
  primaryClass  = {math.DS}
}

@article{treschev1996averaging,
  author  = {Treschev, D. V.},
  title   = {An averaging method for {H}amiltonian systems, exponentially close to integrable ones},
  journal = {Chaos},
  volume  = {6},
  number  = {1},
  pages   = {6--14},
  year    = {1996},
  doi     = {10.1063/1.166149}
}

@article{treschev1997continuous,
  author  = {Treschev, D. V.},
  title   = {The continuous averaging method in the problem of separation of fast and slow motions},
  journal = {Regular and Chaotic Dynamics},
  volume  = {2},
  number  = {3--4},
  pages   = {9--20},
  year    = {1997},
  note    = {In Russian}
}

@inproceedings{treschev2002icm,
  author        = {Treschev, D. V.},
  title         = {Continuous averaging in dynamical systems},
  booktitle     = {Proceedings of the International Congress of Mathematicians, Beijing 2002},
  volume        = {3},
  pages         = {383--394},
  year          = {2002},
  eprint        = {math/0304459},
  archivePrefix = {arXiv}
}

@article{neishtadt1987delay1,
  author  = {Neishtadt, A. I.},
  title   = {Persistence of stability loss for dynamical bifurcations. I},
  journal = {Differential Equations},
  volume  = {23},
  number  = {12},
  pages   = {1385--1391},
  year    = {1987},
  note    = {Translated from Differentsial'nye Uravneniya \textbf{23} (1987), no.~12, 2060--2067}
}

@article{neishtadt1988delay2,
  author  = {Neishtadt, A. I.},
  title   = {Persistence of stability loss for dynamical bifurcations. II},
  journal = {Differential Equations},
  volume  = {24},
  number  = {2},
  pages   = {171--176},
  year    = {1988},
  note    = {Translated from Differentsial'nye Uravneniya \textbf{24} (1988), no.~2, 226--233}
}

@article{benoit1998surstables,
  author  = {Beno{\^i}t, {\'E}ric and Fruchard, Augustin and Sch{\"a}fke, Reinhard and Wallet, Guy},
  title   = {Solutions surstables des {\'e}quations diff{\'e}rentielles complexes lentes-rapides {\`a} point tournant},
  journal = {Ann. Fac. Sci. Toulouse Math. (6)},
  volume  = {7},
  number  = {4},
  pages   = {627--658},
  year    = {1998}
}

@article{fruchard2009overstability,
  author  = {Fruchard, Augustin and Sch{\"a}fke, Reinhard},
  title   = {A survey of some results on overstability and bifurcation delay},
  journal = {Discrete Contin. Dyn. Syst. Ser. S},
  volume  = {2},
  number  = {4},
  pages   = {931--965},
  year    = {2009},
  doi     = {10.3934/dcdss.2009.2.931}
}

@article{canalisdurand2000gevrey,
  author  = {Canalis-Durand, Mireille and Ramis, Jean-Pierre and Sch{\"a}fke, Reinhard and Sibuya, Yasutaka},
  title   = {Gevrey solutions of singularly perturbed differential equations},
  journal = {J. Reine Angew. Math.},
  volume  = {518},
  pages   = {95--129},
  year    = {2000},
  doi     = {10.1515/crll.2000.008}
}

@article{benoit2002combined,
  author  = {Benoit, Eric and El Hamidi, Abdallah and Fruchard, Augustin},
  title   = {On combined asymptotic expansions in singular perturbations},
  journal = {Electronic Journal of Differential Equations},
  volume  = {2002},
  number  = {51},
  pages   = {1--27},
  year    = {2002}
}

@book{fruchard2013composite,
  author    = {Fruchard, Augustin and Sch{\"a}fke, Reinhard},
  title     = {Composite Asymptotic Expansions},
  series    = {Lecture Notes in Mathematics},
  volume    = {2066},
  pages     = {x+161},
  year      = {2013},
  publisher = {Springer},
  address   = {Berlin, Heidelberg},
  doi       = {10.1007/978-3-642-34035-2},
  isbn      = {978-3-642-34034-5}
}

@misc{nikolaev2022existence,
  author        = {Nikolaev, Nikita},
  title         = {An Exact Perturbative Existence and Uniqueness Theorem},
  year          = {2022},
  doi           = {10.48550/arXiv.2201.04526},
  eprint        = {2201.04526},
  archivePrefix = {arXiv},
  primaryClass  = {math.CA}
}

@article{nikolaev2024gevrey,
  author        = {Nikolaev, Nikita},
  title         = {Gevrey Asymptotic Implicit Function Theorem},
  journal       = {L'Enseignement Math\'{e}matique},
  volume        = {70},
  pages         = {251--282},
  year          = {2024},
  doi           = {10.4171/LEM/1061},
  eprint        = {2112.08792},
  archivePrefix = {arXiv},
  primaryClass  = {math.CV}
}

@incollection{ecalle1993transseries,
  author    = {{\'E}calle, Jean},
  title     = {Six lectures on transseries, analysable functions and the constructive proof of Dulac's conjecture},
  booktitle = {Bifurcations and periodic orbits of vector fields},
  pages     = {75--184},
  year      = {1993},
  publisher = {Springer},
  doi       = {10.1007/978-94-015-8238-4_3}
}

@article{basar2017quantumgeometry,
  author        = {Ba{\c{s}}ar, G{\"o}k{\c{c}}e and Dunne, Gerald V. and {\"U}nsal, Mithat},
  title         = {Quantum geometry of resurgent perturbative/nonperturbative relations},
  journal       = {Journal of High Energy Physics},
  volume        = {2017},
  number        = {5},
  pages         = {087},
  year          = {2017},
  publisher     = {Springer},
  doi           = {10.1007/JHEP05(2017)087},
  eprint        = {1701.06572},
  archivePrefix = {arXiv},
  primaryClass  = {hep-th}
}

@inproceedings{dunne2017mathieu,
  author       = {Dunne, Gerald V. and {\"U}nsal, Mithat},
  title        = {{WKB} and resurgence in the {M}athieu equation},
  booktitle    = {Resurgence, physics and numbers},
  pages        = {249--298},
  year         = {2017},
  doi          = {10.1007/978-88-7642-613-1_6},
  organization = {Springer}
}

@article{aniceto2019primer,
  author    = {Aniceto, In{\^e}s and Ba{\c{s}}ar, G{\"o}k{\c{c}}e and Schiappa, Ricardo},
  title     = {A primer on resurgent transseries and their asymptotics},
  journal   = {Physics Reports},
  volume    = {809},
  pages     = {1--135},
  year      = {2019},
  publisher = {Elsevier},
  doi       = {10.1016/j.physrep.2019.02.003}
}

@article{dorigoni2019introduction,
  author    = {Dorigoni, Daniele},
  title     = {An introduction to resurgence, trans-series and alien calculus},
  journal   = {Annals of Physics},
  volume    = {409},
  pages     = {167914},
  year      = {2019},
  publisher = {Elsevier},
  doi       = {10.1016/j.aop.2019.167914}
}

@book{kawai2005algebraic,
  author    = {Kawai, Takahiro and Takei, Yoshitsugu},
  title     = {Algebraic Analysis of Singular Perturbation Theory},
  series    = {Translations of Mathematical Monographs},
  volume    = {227},
  year      = {2005},
  publisher = {American Mathematical Society},
  address   = {Providence, RI},
  isbn      = {978-0-8218-3547-0}
}

@article{sueishi2020quantization,
  author    = {Sueishi, Naohisa and Kamata, Syo and Misumi, Tatsuhiro and {\"U}nsal, Mithat},
  title     = {On exact-WKB analysis, resurgent structure, and quantization conditions},
  journal   = {Journal of High Energy Physics},
  volume    = {2020},
  number    = {12},
  pages     = {114},
  year      = {2020},
  publisher = {Springer},
  doi       = {10.1007/JHEP12(2020)114}
}

@article{sueishi2021anomaly,
  author    = {Sueishi, Naohisa and Kamata, Syo and Misumi, Tatsuhiro and {\"U}nsal, Mithat},
  title     = {Exact-WKB, complete resurgent structure, and mixed anomaly in quantum mechanics on $S^1$},
  journal   = {Journal of High Energy Physics},
  volume    = {2021},
  number    = {7},
  pages     = {96},
  year      = {2021},
  publisher = {Springer},
  doi       = {10.1007/JHEP07(2021)096}
}

@article{nikolaev2023riccati,
  author        = {Nikolaev, Nikita},
  title         = {Exact Solutions for the Singularly Perturbed {R}iccati Equation and Exact {WKB} Analysis},
  journal       = {Nagoya Mathematical Journal},
  volume        = {250},
  pages         = {434--469},
  year          = {2023},
  doi           = {10.1017/nmj.2022.38},
  eprint        = {2008.06492},
  archivePrefix = {arXiv},
  primaryClass  = {math.CA}
}

@article{nikolaev2023exactwkb,
  author        = {Nikolaev, Nikita},
  title         = {Existence and Uniqueness of Exact {WKB} Solutions for Second-Order Singularly Perturbed Linear {ODE}s},
  journal       = {Communications in Mathematical Physics},
  volume        = {400},
  number        = {1},
  pages         = {463--517},
  year          = {2023},
  doi           = {10.1007/s00220-022-04603-7},
  eprint        = {2106.10248},
  archivePrefix = {arXiv},
  primaryClass  = {math.AP}
}

@article{kristiansen2024wkbturning,
  author    = {Kristiansen, K. U. and Szmolyan, P.},
  title     = {A dynamical systems approach to {{WKB}}-methods: {T}he simple turning point},
  journal   = {Journal of Differential Equations},
  volume    = {406},
  pages     = {202--254},
  year      = {2024},
  publisher = {Elsevier},
  doi       = {10.1016/j.jde.2024.06.006}
}

@misc{nikolaev2024geometry,
  author        = {Nikolaev, Nikita},
  title         = {Geometry and Resurgence of {WKB} Solutions of {S}chr\"{o}dinger Equations},
  year          = {2024},
  doi           = {10.48550/arXiv.2410.17224},
  eprint        = {2410.17224},
  archivePrefix = {arXiv},
  primaryClass  = {math.DG}
}

@article{kristiansen2025wkbeigenvalue,
  author        = {Kristiansen, K. U. and Szmolyan, P.},
  title         = {A dynamical systems approach to {WKB}-methods: {T}he eigenvalue problem for a single well potential},
  journal       = {Studies in Applied Mathematics},
  volume        = {155},
  number        = {5},
  pages         = {e70141},
  year          = {2025},
  doi           = {10.1111/sapm.70141},
  eprint        = {2501.10707},
  archivePrefix = {arXiv},
  primaryClass  = {math-ph}
}

@article{GelfandDikii1975,
    author  = {I. M. Gel'fand and L. A. Dikii},
    title   = {Asymptotic behaviour of the resolvent of {S}turm--{L}iouville equations
    and the algebra of the {K}orteweg--de {V}ries equations},
    journal = {Russian Mathematical Surveys},
    volume  = {30},
    number  = {5},
    pages   = {77--113},
    year    = {1975},
}

@article{Delabaere1993,
    author  = {H. Dillinger and E. Delabaere and F. Pham},
    title   = {R\'esurgence de Voros et p\'eriodes des courbes hyperelliptiques},
    journal = {Annales de l'Institut Fourier},
    volume  = {43},
    number  = {1},
    pages   = {163--199},
    year    = {1993},
}

@article{Delabaere1997,
    author  = {E. Delabaere and H. Dillinger and F. Pham},
    title   = {Exact semiclassical expansions for one-dimensional quantum oscillators},
    journal = {Journal of Mathematical Physics},
    volume  = {38},
    number  = {12},
    pages   = {6126--6184},
    year    = {1997},
    doi     = {10.1063/1.532206},
}

@book{ByrdFriedman1971,
    author    = {P. F. Byrd and M. D. Friedman},
    title     = {Handbook of Elliptic Integrals for Engineers and Scientists},
    edition   = {2nd, revised},
    series    = {Die Grundlehren der mathematischen Wissenschaften},
    volume    = {67},
    publisher = {Springer-Verlag},
    address   = {Berlin, Heidelberg, New York},
    year      = {1971},
}

@article{renne1974shunted,
  author  = {Renne, M. J. and Polder, D.},
  title   = {Some analytical results for the resistively shunted {J}osephson junction},
  journal = {Revue de physique appliqu{\'e}e},
  volume  = {9},
  number  = {1},
  pages   = {25--28},
  year    = {1974},
  doi     = {10.1051/rphysap:019740090102500}
}

@article{guckenheimer2001duckdevil,
  author  = {Guckenheimer, J. and Ilyashenko, Y.},
  title   = {The duck and the devil: canards on the staircase},
  journal = {Moscow Mathematical Journal},
  volume  = {1},
  number  = {1},
  pages   = {27--47},
  year    = {2001},
  doi     = {10.17323/1609-4514-2001-1-1-27-47}
}

@article{buchstaber2010rotation,
  author    = {Buchstaber, V. and Karpov, O. and Tertychniy, S.},
  title     = {Rotation number quantization effect},
  journal   = {Theoretical and Mathematical Physics},
  volume    = {162},
  pages     = {211--221},
  year      = {2010},
  publisher = {Springer},
  doi       = {10.1007/s11232-010-0016-4}
}

@article{schurov2010torus,
  author  = {Shchurov, I.},
  title   = {Canard cycles in generic fast-slow systems on the torus},
  journal = {Transactions of the Moscow Mathematical Society},
  volume  = {71},
  pages   = {175--207},
  year    = {2010},
  doi     = {10.1090/S0077-1554-2010-00184-7}
}

@article{glutsyuk2014adjacency,
  author    = {Glutsyuk, A. and Kleptsyn, V. and Filimonov, D. and Schurov, I.},
  title     = {On the adjacency quantization in an equation modeling the {J}osephson effect},
  journal   = {Functional Analysis and Its Applications},
  volume    = {48},
  number    = {4},
  pages     = {272--285},
  year      = {2014},
  publisher = {Springer},
  doi       = {10.1007/s10688-014-0070-z}
}

@misc{glutsyuk2026parquet,
    title={On phase-lock area parquet in a special slow-fast limit of model of Josephson junction}, 
    author={Alexey Glutsyuk},
    year={2026},
    eprint={2607.26158},
    archivePrefix={arXiv},
    primaryClass={math.DS},
    doi={10.48550/arXiv.2607.26158}
}

@article{schurov2017factory,
  author    = {Schurov, I. and Solodovnikov, N.},
  title     = {Duck factory on the two-torus: multiple canard cycles without geometric constraints},
  journal   = {Journal of Dynamical and Control Systems},
  volume    = {23},
  pages     = {481--498},
  year      = {2017},
  publisher = {Springer},
  doi       = {10.1007/s10883-016-9335-6}
}

@article{glutsyuk2019constrictions,
  author    = {Glutsyuk, A.},
  title     = {On constrictions of phase-lock areas in model of overdamped {J}osephson effect and transition matrix of the double-confluent {H}eun equation},
  journal   = {Journal of Dynamical and Control Systems},
  volume    = {25},
  number    = {3},
  pages     = {323--349},
  year      = {2019},
  publisher = {Springer},
  doi       = {10.1007/s10883-018-9411-1}
}

@article{alexandrov2026phaselocking,
  author    = {Alexandrov, Artem and Glutsyuk, Alexey and Gorsky, Alexander},
  title     = {Phase-locking in dynamical systems and quantum mechanics},
  journal   = {Journal of High Energy Physics},
  volume    = {2026},
  number    = {3},
  pages     = {101},
  year      = {2026},
  publisher = {Springer Berlin Heidelberg},
  doi       = {10.1007/JHEP03(2026)101}
}

@article{Shapiro1963,
  author = {S. Shapiro},
  title = {Josephson Currents in Superconducting Tunneling:
  The Effect of Microwaves and Other Observations},
  journal = {Physical Review Letters},
  volume = {11},
  number = {2},
  pages = {80--82},
  year = {1963},
  doi = {10.1103/PhysRevLett.11.80}
}

@book{Likharev1986,
  author = {K. K. Likharev},
  title = {Dynamics of Josephson Junctions and Circuits},
  publisher = {Gordon and Breach},
  address = {New York},
  year = {1986}
}

@book{BaronePaterno1982,
  author = {A. Barone and G. Patern\`o},
  title = {Physics and Applications of the
  Josephson Effect},
  publisher = {Wiley},
  address = {New York},
  year = {1982},
  doi = {10.1002/352760278X}
}

@article{gandhi2015dynamics,
  doi={PhysRevE.92.062914},
  title={Dynamics of phase slips in systems with time-periodic modulation},
  author={Gandhi, Punit and Knobloch, Edgar and Beaume, C{\'e}dric},
  journal={Physical Review E},
  volume={92},
  number={6},
  pages={062914},
  year={2015},
  publisher={APS}
}

@article{varadhan1966asymptotic,
  author  = {Varadhan, S. R. S.},
  title   = {Asymptotic probabilities and differential equations},
  journal = {Communications on Pure and Applied Mathematics},
  volume  = {19},
  number  = {3},
  pages   = {261--286},
  year    = {1966},
  doi     = {10.1002/cpa.3160190303}
}

@article{sowers2008noisycanards,
  author  = {Sowers, Richard B.},
  title   = {Random Perturbations of Canards},
  journal = {Journal of Theoretical Probability},
  volume  = {21},
  number  = {4},
  pages   = {824--889},
  year    = {2008},
  doi     = {10.1007/s10959-008-0150-1}
}

@article{touchette2009largedeviation,
  author  = {Touchette, Hugo},
  title   = {The large deviation approach to statistical mechanics},
  journal = {Physics Reports},
  volume  = {478},
  number  = {1--3},
  pages   = {1--69},
  year    = {2009},
  doi     = {10.1016/j.physrep.2009.05.002}
}

@book{dembo2010largedeviations,
  author    = {Dembo, Amir and Zeitouni, Ofer},
  title     = {Large Deviations Techniques and Applications},
  series    = {Stochastic Modelling and Applied Probability},
  volume    = {38},
  year      = {2010},
  publisher = {Springer},
  address   = {Berlin, Heidelberg},
  edition   = {2},
  doi       = {10.1007/978-3-642-03311-7}
}

@article{berglund2012noisy,
  author    = {Berglund, Nils and Gentz, Barbara and Kuehn, Christian},
  title     = {Hunting French ducks in a noisy environment},
  journal   = {Journal of Differential Equations},
  volume    = {252},
  number    = {9},
  pages     = {4786--4841},
  year      = {2012},
  publisher = {Elsevier},
  doi       = {10.1016/j.jde.2012.01.015}
}

@book{freidlin2012random,
  author    = {Freidlin, Mark I. and Wentzell, Alexander D.},
  title     = {Random Perturbations of Dynamical Systems},
  series    = {Grundlehren der mathematischen Wissenschaften},
  volume    = {260},
  year      = {2012},
  publisher = {Springer},
  address   = {Berlin, Heidelberg},
  edition   = {3},
  doi       = {10.1007/978-3-642-25847-3}
}

@inproceedings{prandtl1904boundarylayer,
  author    = {Prandtl, Ludwig},
  title     = {{\"U}ber Fl{\"u}ssigkeitsbewegung bei sehr kleiner Reibung},
  booktitle = {Verhandlungen des III. Internationalen Mathematiker-Kongresses, Heidelberg 1904},
  pages     = {484--491},
  year      = {1905},
  publisher = {Teubner},
  address   = {Leipzig}
}

@article{chen1994rg,
  author  = {Chen, Lin-Yuan and Goldenfeld, Nigel and Oono, Y.},
  title   = {Renormalization group theory for global asymptotic analysis},
  journal = {Physical Review Letters},
  volume  = {73},
  number  = {10},
  pages   = {1311--1315},
  year    = {1994},
  doi     = {10.1103/PhysRevLett.73.1311}
}

@article{chen1996rg,
  author  = {Chen, Lin-Yuan and Goldenfeld, Nigel and Oono, Y.},
  title   = {Renormalization group and singular perturbations: Multiple scales, boundary layers, and reductive perturbation theory},
  journal = {Physical Review E},
  volume  = {54},
  number  = {1},
  pages   = {376--394},
  year    = {1996},
  doi     = {10.1103/PhysRevE.54.376}
}

@book{schlichting2017boundarylayer,
  author    = {Schlichting, Hermann and Gersten, Klaus},
  title     = {Boundary-Layer Theory},
  year      = {2017},
  publisher = {Springer},
  address   = {Berlin, Heidelberg},
  edition   = {9},
  doi       = {10.1007/978-3-662-52919-5}
}

@article{panghotra2020giant,
  doi={10.1038/s42005-020-0315-5},
  title={Giant fractional {S}hapiro steps in anisotropic {J}osephson junction arrays},
  author={Panghotra, R. and Raes, B. and de Souza Silva, C. C. and Cools, I. and Keijers, W. and Scheerder, J. E. and Moshchalkov, V. V. and Van de Vondel, J.},
  journal={Communications Physics},
  volume={3},
  number={1},
  pages={53},
  year={2020},
  publisher={Nature Publishing Group UK London}
}

@article{stolyarov2026shapiro,
  title={Shapiro steps in ballistic Josephson junction based on a single Bi2Te2. 3Se0. 7 nanocrystal},
  author={Stolyarov, VS and Kozlov, SN and Yakovlev, DS and Skryabina, OV and Lvov, DS and Vasenko, AS and Zhou, J and Kupriyanov, M Yu and Golubov, AA and Feuillet-Palma, C and others},
  journal={Communications Materials},
  volume={7},
  number={1},
  pages={91},
  year={2026},
  publisher={Nature Publishing Group UK London},
  doi={10.1038/s43246-026-01095-z}
}

@article{brons1991belousov,
  author  = {Br{\o}ns, M. and Bar-Eli, K.},
  title   = {Canard explosion and excitation in a model of the {B}elousov--{Z}habotinskii reaction},
  journal = {The Journal of Physical Chemistry},
  volume  = {95},
  number  = {22},
  pages   = {8706--8713},
  year    = {1991},
  doi     = {10.1021/j100175a053}
}

@article{moehlis2006hodgkinhuxley,
  author  = {Moehlis, Jeff},
  title   = {Canards for a reduction of the {H}odgkin--{H}uxley equations},
  journal = {Journal of Mathematical Biology},
  volume  = {52},
  number  = {2},
  pages   = {141--153},
  year    = {2006},
  doi     = {10.1007/s00285-005-0347-1}
}

@article{rotstein2008stellate,
  author  = {Rotstein, Horacio G. and Wechselberger, Martin and Kopell, Nancy},
  title   = {Canard Induced Mixed-Mode Oscillations in a Medial Entorhinal Cortex Layer {II} Stellate Cell Model},
  journal = {SIAM Journal on Applied Dynamical Systems},
  volume  = {7},
  number  = {4},
  pages   = {1582--1611},
  year    = {2008},
  doi     = {10.1137/070706778}
}

@article{harvey2011calcium,
  author  = {Harvey, Emily and Kirk, Vivien and Wechselberger, Martin and Sneyd, James},
  title   = {Multiple Timescales, Mixed Mode Oscillations and Canards in Models of Intracellular Calcium Dynamics},
  journal = {Journal of Nonlinear Science},
  volume  = {21},
  number  = {5},
  pages   = {639--683},
  year    = {2011},
  doi     = {10.1007/s00332-011-9096-z}
}

@article{desroches2013bursting,
  author  = {Desroches, Mathieu and Kaper, Tasso J. and Krupa, Martin},
  title   = {Mixed-Mode Bursting Oscillations: Dynamics Created by a Slow Passage Through Spike-Adding Canard Explosion in a Square-Wave Burster},
  journal = {Chaos},
  volume  = {23},
  number  = {4},
  pages   = {046106},
  year    = {2013},
  doi     = {10.1063/1.4827026}
}

@article{vo2013pseudoplateau,
  author  = {Vo, Theodore and Bertram, Richard and Wechselberger, Martin},
  title   = {Multiple Geometric Viewpoints of Mixed Mode Dynamics Associated with Pseudo-Plateau Bursting},
  journal = {SIAM Journal on Applied Dynamical Systems},
  volume  = {12},
  number  = {2},
  pages   = {789--830},
  year    = {2013},
  doi     = {10.1137/120892842}
}

@article{ginoux2016memristor,
  author  = {Ginoux, Jean-Marc and Llibre, Jaume},
  title   = {Canards Existence in Memristor's Circuits},
  journal = {Qualitative Theory of Dynamical Systems},
  volume  = {15},
  number  = {2},
  pages   = {383--431},
  year    = {2016},
  doi     = {10.1007/s12346-015-0160-1}
}

@article{zhan2018pituitary,
  author  = {Zhan, Feibiao and Liu, Shenquan and Zhang, Xiaohan and Wang, Jing and Lu, Bo},
  title   = {Mixed-Mode Oscillations and Bifurcation Analysis in a Pituitary Model},
  journal = {Nonlinear Dynamics},
  volume  = {94},
  number  = {2},
  pages   = {807--826},
  year    = {2018},
  doi     = {10.1007/s11071-018-4395-7}
}

@article{liu2020pyramidal,
  author  = {Liu, Yaru and Liu, Shenquan},
  title   = {Canard-Induced Mixed-Mode Oscillations and Bifurcation Analysis in a Reduced {3D} Pyramidal Cell Model},
  journal = {Nonlinear Dynamics},
  volume  = {101},
  number  = {1},
  pages   = {531--567},
  year    = {2020},
  doi     = {10.1007/s11071-020-05801-5}
}

\end{document}